\pdfoutput=1
\documentclass[10pt]{article}
\usepackage[applemac]{inputenc}
\usepackage{graphicx}
\usepackage{amssymb}
\usepackage{url}
\newcommand{\dd}{{\rm d}}

\newcommand{\bb}{\mathbf}

\begin{document}

\title{2D electromagnetic simulations of RF heating via inductive coupling in the SPIDER device}
\author{D. L\'opez-Bruna$^1$, P. Jain$^2$, M. Recchia$^2$, B. Zaniol$^2$, E. Sartori$^2$,  \\C. Poggi$^2$, V. Candeloro$^2$, G. Serianni$^2$, P. Veltri$^3$}
\date{\footnotesize 
\begin{flushleft}
$^1$ Laboratorio Nacional de Fusi\'on, CIEMAT, Avda.~Complutense 40 28040 Madrid, Spain \\
$^2$ Consorzio RFX (CNR, ENEA, INFN, UNIPD, Acciaierie Venete SpA), Corso Stati Uniti 4 35127 Padova, Italy \\
$^3$ ITER Organization, Route de Vinon-sur-Verdon, CS 90 046, F-13067 St. Paul-lez-Durance, France\\
\end{flushleft}(17 February 2022)}

\maketitle

\begin{abstract}
{\small SPIDER is the prototype ion source of MITICA, the full-size neutral beam heating system conceived for the ITER tokamak. It includes eight drivers to heat and sustain the inductively coupled plasma (ICP). Owing to their near cylindrical symmetry, the coupling between the radio-frequency (RF) active currents and the source plasma is studied using a 2D electromagnetic approach with simplified expressions for the plasma electrical conductivity taken from the literature. The power absorbed by the plasma and the effect of the induced plasma currents in lowering the inductance of the driver are based on data from the dedicated S16 experimental campaign (y.~2020) of SPIDER: plasma electron densities on the order of $10^{18}$ m$^{-3}$, electron temperatures $\sim 10$ eV; neutral gas pressure $\sim 0.3$ Pa and up to $50$ kW of net power per driver. It is found that the plasma conductivity cannot be explained by the friction forces associated to local collisional processes alone. The inclusion of an effective collisionality associated to non-local processes seems also insufficient to explain the experimental information. Only when the electrical conductivity is reduced where the RF magnetic field is more intense, can the heating power and driver inductance be acceptably reproduced. We present the first 2D electromagnetic ICP calculations in SPIDER for two types of plasma, without and with the addition of a static magnetic field. The power transfer efficiency to the plasma of the first drivers of SPIDER, in view of these models, is around $50$\%.}
\end{abstract}

\section{Introduction}

The heating and sustainment of the plasma in the source of the SPIDER (Source for the Production of Ions of Deuterium Extracted from a Radio frequency plasma) device \cite{Serianni2019SPIDER-in-the-r,Serianni2020First-operation,Toigo2017The-PRIMA-Test-} is realized by means of eight  heaters based on the inductive coupling between radio frequency waves and the plasma (ICP).  Each heater, commonly called ``driver'', consists of a cylindrical chamber with one open end facing the plasma expansion region. Here, the hot plasma generated in the driver diffuses and its temperature decreases so as to enhance the survival probability of negative ions prior to their extraction into an energetic beam. The ICP is, consequently, an important element in the simulation of the plasma in the driver and expansion regions.
 Previous works in SPIDER have been dedicated to obtain numerical tools to estimate the fraction of the net input power that is absorbed by the plasma, namely the power transfer efficiency, depending on main external controllers of the drivers like the power output from the generator, the value of the RF and the gas pressure in the chamber of the drivers \cite{Jain2018Improved-Method,Recchia2021studies-on-powe}. In this case, a basic zero-dimensional transport problem is solved in order to provide some meaningful feedback between the power delivered to the plasma and its characteristics, which in turn determine the induction process through the plasma conductivity. With the present paper we start a program of complementary calculations of the inductive coupling between the active RF current and the plasma considering spatial distributions of electron density and temperature. As a first step, we solve the equation for the induced electric field in the plasma region assuming perfect cylindrical symmetry, which renders the problem bi-dimensional (2D). Since one important objective is to provide a heating module for the available 2D fluid transport code \cite{zagorski20222-d-fluid-model}, we impose 2D plasmas with parameters taken from the experimental campaigns of the SPIDER device without caring for the transport problem.

Electromagnetic calculations are based on well-known equations. Once the boundary conditions are established, the essential difficulty in order to gain information out of the calculations consists of the properties of the medium, i.e., the conductivity of the plasma inside the driver. This is a theoretical problem. On the other hand, the possibility to confront the consequences of a model conductivity with the experimental data requires that the thermodynamic properties of the plasma be known to an acceptable extent. This is an experimental problem. There are many works devoted to the complicated theoretical problem (see \cite{chabert2021foundations-of-} and references therein), which is out of the scope of this paper. On the other hand, previous works for SPIDER have made use of simplified expressions for the plasma conductivity that seem to give a fair account of the induction process. The development of reliable calculation tools is a mandatory step before the physics of electrical conduction can be tackled. Therefore, in this work we use experimental information to study the ICP according to the same electrical conductivity models in order to check their adequacy to represent the behaviour of SPIDER plasmas, thus setting the basis for further studies.

After this introduction, Section \ref{sec:2} gives a brief description of SPIDER and the 2D electromagnetic calculations.
  In Section \ref{sec:datos experimentales} we present the experimental information that will be used later on, which concerns the basic experimental knobs (gas pressure and species, current in the RF coils,  RF frequency and delivered power) and the distributions of electron density and temperature inside the drivers.
   Section \ref{sec:conductividad} is dedicated to briefly recall the models of plasma conductivity that will be used in the calculations presented in Section \ref{sec:resultados}. The work finishes with a discussion (Section \ref{sec:discusion}) based on calculations of the absorbed power in plasmas without and with filter magnetic field, and a summary (Section \ref{sec:conclusion}).
      Details of the induction equation are given in Appendix \ref{apendice:ecuaciones}.


\section{2D electromagnetic model of the drivers\label{sec:2}}

\subsection{Basic geometry and parameters}

\begin{figure}
   \centering
   \includegraphics[width=0.4\columnwidth]{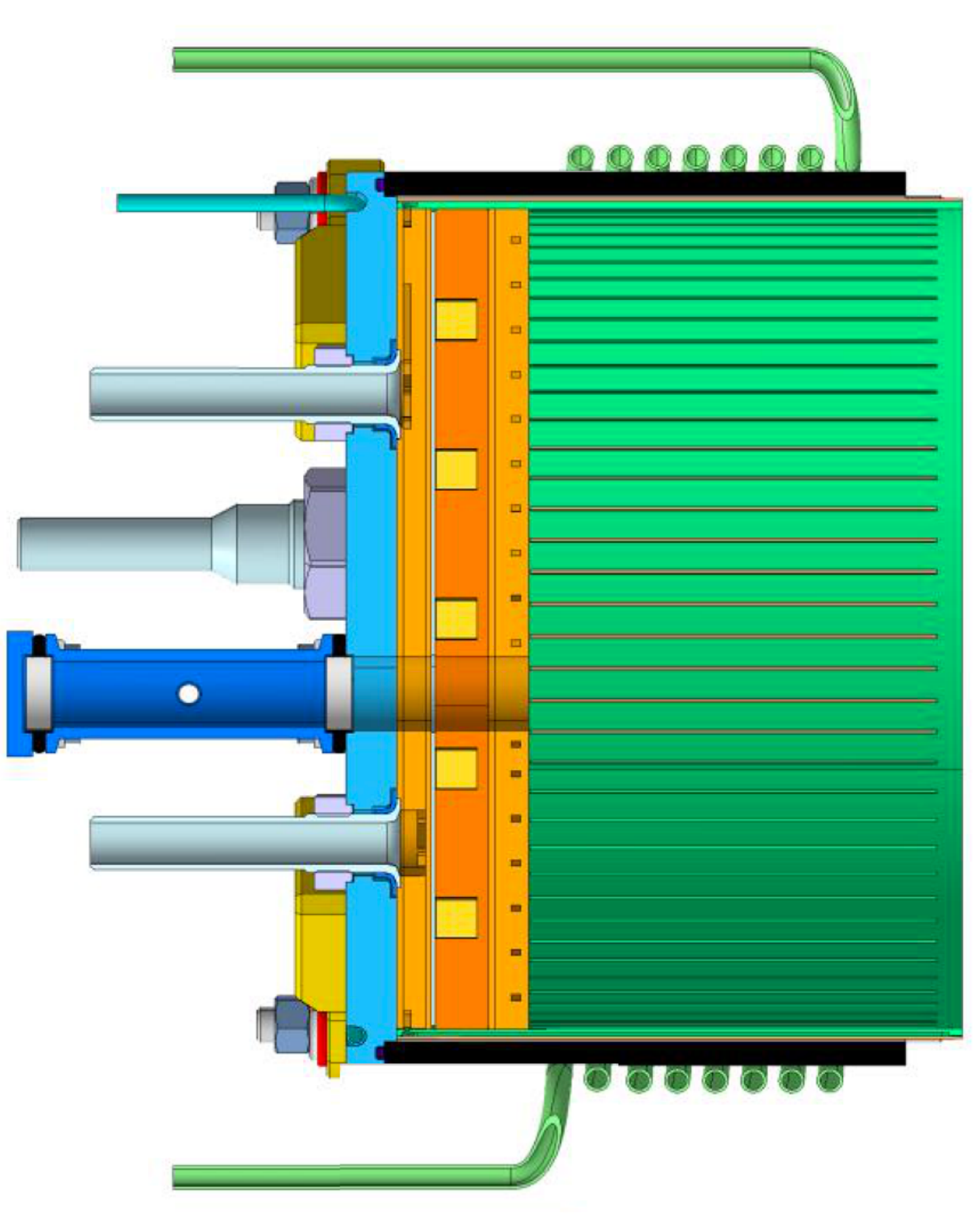}
    \caption{Side view if a driver of the SPIDER device showing the gas inlets and rear plate at the left side, the lateral Faraday shield structure that covers the plasma chamber (in green) and the section of the RF winding.}
   \label{fig:cazuela_vistas}
\end{figure}

The plasma source in SPIDER is based on the prototype ELISE developed at IPP Garching \cite{heinemann2011the-negative-io}. It is composed of four pairs of drivers of circular cross-section designed to operate at 200 kW to give a total maximum power of 800 kW. Figure \ref{fig:cazuela_vistas} shows a schematic view of one of the eight identical drivers. The filling gas (H$_2$ or D$_2$) is injected through the rear plate inlets at a pressure of  $\sim 0.3$ Pa and is heated mainly using induction coils wound around each driver. The coil is separated from the driver chamber by an alumina insulator that covers the cylindrical side of the Faraday shield. The latter is a copper structure that protects the insulator from sputtering, while permitting the RF field penetration in the plasma region through eighty shaped incisions. The RF-sustained plasma diffuses out to the expansion chamber and towards the plasma grid (PG) surface, where a Cesium coating decreases the work function in order to favour the creation of negative ions \cite{toigo2021on-the-road-to-}. Just before the PG there is a biased plate divided in five segments. Its adjustable potential allows for absorbing variable quantities of electrons with the aim of optimizing the electric field distribution in the region where the negative ions must be produced. See Refs.~\cite{Serianni2019SPIDER-in-the-r,Toigo2019Progress-in-the,Serianni2020First-operation} for more details about the experimental device.

Due to technical reasons, the power delivered by each generator during the 2020 campaigns was limited to 100 kW, about half the design value; i.e., the power was limited to 50 kW per driver. Table \ref{T:parametros iniciales Roman} collects the main parameters associated to the drivers of SPIDER.
In next campaigns the precise geometry will be slightly different to the one used so far: the driver case will be made of quartz with a reduced spacing between RF winding and driver \cite{maistrello2021improvements-in}. In addition, the coil will make exactly eight turns around the driver, instead of the present $8.5$ turns. At present, in consideration of the comparison with data of the 2020 campaigns, we use the values in Table \ref{T:parametros iniciales Roman} and define a model driver approximating its geometry by a cylinder sector in a radial--axial plane of dimensions $r_\mathrm{p}\times \Delta z_\mathrm{p}$. The RF winding is simplified to a set of eight filamentary circular wires located $1.7$ cm away (in radius) of the plasma region and centered along the axis.  As we shall see, a modification of the vacuum field created by these wires can reasonably describe the induction process.  The frequency is fixed to $f=1$ MHz.

\begin{table}
\caption{Main parameters of the drivers in the SPIDER device.}
{\small 
\begin{center}
\begin{tabular}{ll}
driver chamber (plasma) radius & $r_\mathrm{p}=0.137$ m \\
driver depth & $\Delta z_\mathrm{p}\equiv z_\mathrm{end}-z_\mathrm{ini} =0.149$ m \\
RF winding width & $W_\mathrm{b}=0.096$ m \\
RF winding radius & $R_\mathrm{b}=0.154$ m \\
feeding current amplitude & $I_\mathrm{RF}\sim 200$ A \\
feeding current frequency & $f \approx 10^6$ Hz \\
power & $10$--$100$ kW \\
driver vacuum impedance & $R_\mathrm{d} \sim 2$ $\Omega$; $L_\mathrm{d}\sim 10$ $\mu$H \\
electron temperature  & $T_e \sim 10$ eV \\
electron density & $n_e \sim  10^{18}$ m$^{-3}$ \\
gas pressure (H$_2$) & $\sim 0.3$ Pa \\
Filter magnetic field & $0 \sim 4$ mT \\
\end{tabular}
\end{center}
}
\label{T:parametros iniciales Roman}
\end{table}%

The design of SPIDER includes controllable currents in the PG and nearby conducting bus-bars to produce a static filter magnetic field, which is dedicated to reduce both the temperature and the amount of plasma electrons just before the PG. In principle, the filter field is designed to act mainly in the expansion region, but there is still a non negligible stray magnetic field inside the drivers, especially near the opening to the expansion region. The design of the filter-field circuitry has been improved so as to make the field approximately perpendicular to the axis of the cylindrical cavity of the drivers. The values of the filter magnetic field near the exit of the driver (the opening to the expansion region) is not identical for all drivers because there is some dependence on their vertical position, being somewhat more intense at the top and bottom of the drivers layout. The magnetic field intensity decreases towards the driver interior down to $\approx 2/3$ of the value at the exit \cite{marconato2021an-optimized-an}. The operation of the SPIDER source is often done with PG currents on the order of the kA, producing fields in the mT range.
 Experiments have also been done without the filter field, which provide the reference plasmas to study magnetic field effects. At the same time, even in the absence of filter field, the amplitude of the induced magnetic fields can easily reach values of several mT in the main volume of the drivers.

\subsection{Electromagnetic model}

\subsubsection{\label{ecs y cc contorno}Equation and boundary conditions}

Given the relatively low frequency of the RF feeding currents and the fact that we are solving for the induced electric field $\bb{E}$ in the plasma domain, where there are no active currents, the electromagnetic problem reduces to solving the homogeneous equation
\begin{equation}
\nabla^2 \mathbf{E} - \imath \omega \mu_0 \sigma\mathbf{E} = 0.
\label{ec:ecuacion em}
\end{equation}
 The angular frequency $\omega$ comes from considering the current in the RF winding as a single harmonic, $I_\mathrm{RF} \propto \cos ( \omega t )$. Aside from geometrical data, the input for the calculations consists of the amplitude of this current and the characteristics of the medium, which in this work is represented by a scalar conductivity, $\sigma$. The assumed cylindrical symmetry of the problem allows us to reduce the three equations associated with $\bb{E}$ to one equation for the azimuthal component $E_\theta$.  \ref{apendice:ecuaciones} gives the details.

Equation \ref{ec:ecuacion em} represents a typical boundary-value problem. The solution in vacuum is known for a filamentary circular loop (see, for instance, Refs.~\cite{Silvester1968Modern-Electrom,Jackson1998Classical-elect}) and can be obtained numerically as a function of the elliptic integrals $K$ and $E$,
\begin{equation}
G(k) = \frac{(2-k^2)K(k^2) - 2E(k^2)}{k},
\label{ec:lido23_2}
\end{equation}
with the variable 
$$
k^2 = \frac{4R_\mathrm{b}r}{(R_\mathrm{b}+r)^2 + (z-z_\mathrm{b})^2}.
$$
Here $R_\mathrm{b}$ and $z_\mathrm{b}$ represent, respectively, the radius from the axis and the constant-$z$ plane where a given loop lies. 
 We note that $K(0)=E(0)$ and the function $G$ at the axis ($r=0 \Rightarrow k^2=0$) is proportional to $K(0)-E(0)$. The corresponding limit \cite{abramowitz1965handbook}  is
\begin{equation}
\lim_{k \to 0} G(0) = \lim_{k\to 0}2k \frac{K(0)-E(0)}{k^2} = \lim_{k \to 0} 2k\frac{\pi}{4} = 0.
\label{ec:lido23_4}
\end{equation}
With these functions, the vector potential produced in space by one filamentary circular coil of radius $R_\mathrm{b}$ carrying a current $I_\mathrm{RF}$ is
\begin{equation}
A_\theta(r,z) = \frac{\mu_0 I_\mathrm{RF}}{2\pi}\sqrt{\frac{R_\mathrm{b}}{r}}G(k).
\label{ec:silvester}
\end{equation}
Since $E_\theta = -\imath \omega A_\theta$, the vacuum field in presence of $N_\mathrm{b}$ current loops of radius $R_{\mathrm{b}_i}$ and centered in $z_{\mathrm{b}_i}$ is the addition of the fields produced by each of them,
\begin{equation}
E_\theta^\mathrm{V}(r,z) = -\imath f \mu_0 \sum_{i=1}^{N_\mathrm{b}} I_{\mathrm{b}_i}\sqrt{\frac{R_{\mathrm{b}_i}}{r}} G(r,z,R_{\mathrm{b}_i},z_{\mathrm{b}i}),
\label{ec:campoEthetav}
\end{equation}
where $f=\omega/2\pi$. The radii $R_{\mathrm{b}_i}=R_\mathrm{b}$ and the currents $I_{\mathrm{b}_i}=I_\mathrm{RF}$ are common to the loops of the RF winding with which we describe the true spiral winding in figure \ref{fig:cazuela_vistas}, while the values $z_{\mathrm{b}i}$ change for each of the $N_\mathrm{b}$ loops. According to the design of the drivers in SPIDER, we take  $N_\mathrm{b}=8$.  
 Expressions \ref{ec:silvester} and \ref{ec:campoEthetav} are valid due to the relatively small $\omega$ and the small dimensions of the driver, negligible with respect to the RF wavelength.

\begin{figure}[htbp] 
    \centering
       (a)\includegraphics[width=0.43\columnwidth]{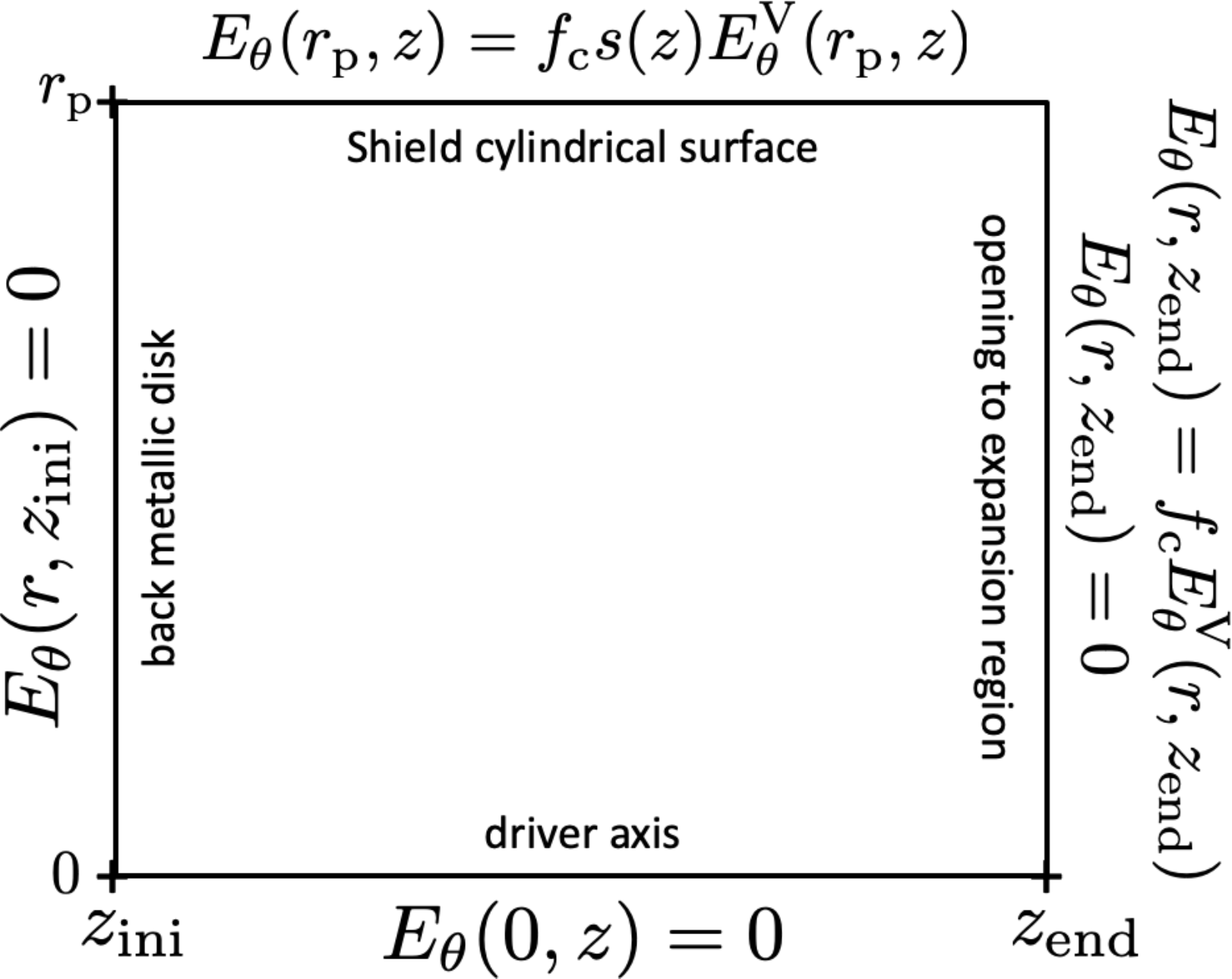} 
       (b)\includegraphics[width=0.38\columnwidth]{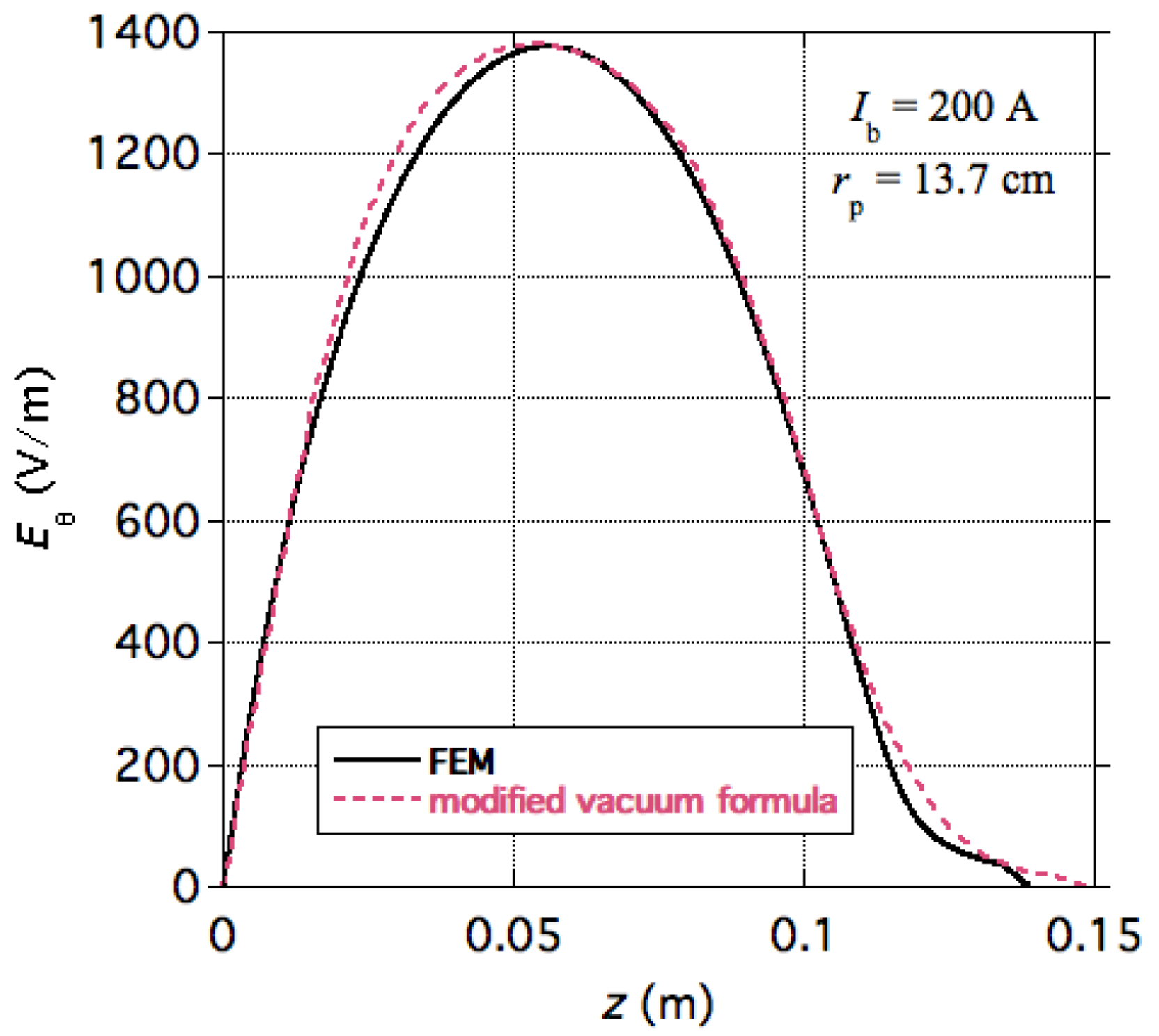} 
    \caption{(a) Boundary conditions at the four sides of the rectangular grid used in the induced electric field calculations. Two options are showed for the opening to the expansion region. (b) Imaginary part of the induced electric field without plasma from a finite-element-method calculation that includes passive elements (black line) and from the modified Eq.~\ref{eq:void} for the vacuum theoretical field, Eq. \ref{ec:campoEthetav} (red dashes).}
   \label{fig:conds_contorno}
 \end{figure}

 Strictly speaking, the boundary conditions should be taken far enough from the system, where it can be said that the fields are zero to good  approximation, and then solve for all the current densities in the calculation domain including all passive conductors and the RF winding itself. In order to allow for a calculation in the plasma domain alone, appropriate boundary conditions must be found. There are four segments where boundary conditions apply, see figure \ref{fig:conds_contorno}a. At $r=0$ (driver axis) we take the condition $E_\theta = 0$ pertinent to the cylindrical symmetry of the problem. At the back side of the driver, $z = z_\mathrm{ini}$, there is a molybdenum-coated copper disk. The large difference between the conductivities in the metalic disk and the plasma, together with the known condition of equal field components tangent to the boundary, allows us to set the field in this boundary to zero. There are two remaining boundaries,  the side facing the expansion zone and the line at the cylindrical surface of the Faraday shield. The theoretical solution in vacuum,  Eq.~\ref{ec:campoEthetav}, is useful to obtain adapted boundary conditions in the absence of plasma. In the presence of plasma, however, the currents induced in it will further modify the boundary values, which calls for an iterative process.

\paragraph{Void driver}

Let us begin with the boundary conditions for a void driver, i.e., for the driver without plasma but considering the passive metallic structures.  We start by considering the vacuum solution, Eq. \ref{ec:campoEthetav}, as a shape function that respects the more intense field near the RF winding; but use a global constant factor $f_\mathrm{c} < 1$ to reduce its maximum value. Since the nature of Eq.~\ref{ec:campoEthetav} imposes $f_\mathrm{c}E_\theta^{\mathrm{V}}(r_\mathrm{p},0) \neq 0$, the condition on the cylindrical lateral surface is forced to zero at $z=z_\mathrm{ini}$ using a smooth step function, $\tanh [(z-z_\mathrm{ini})/2\delta]$, with $\delta = 0.01$ m. Likewise, the field near the opening to the expansion region is very small due to the presence of another metallic casing, which is different from the pure vacuum field due to the circular wires alone. Here again, we set a smooth step function to force near zero values by this metallic border. Together with the former one, we have opted for the following shaping function
\begin{equation}
s(z) = \tanh \left (\frac{z-z_\mathrm{ini}}{2\delta} \right ) \times \frac{1}{2} \left [1+\tanh \left (-\frac{z-0.0417+\delta)}{2\delta} \right ) \right ].
\label{eq:shaping function}
\end{equation}
Figure \ref{fig:conds_contorno}b 
 represents, with a black line, the out-of-phase (imaginary) part of the induced electric field calculated with a 2D Finite-Element-Method for the geometry of the drivers of SPIDER without plasma, where the highly conducting metallic passive parts of the driver are taken into account and the domain includes the RF coils. With red dashes we represent the function
\begin{equation}
E_\theta^\mathrm{void}(r_\mathrm{p},z) = f_\mathrm{c}s(z)E_\theta^{\mathrm{V}}(r_\mathrm{p},z)
\label{eq:void}
\end{equation}
 with $f_\mathrm{c}= 0.38$ to show that our description of the boundary field without plasma at $r=r_\mathrm{p}$ is a reasonable choice. Both calculations have been done using a same current in the winding, which can be taken here as a scaling factor because the problem, so far, is linear. The in-phase (real) part of the field Eq.~\ref{eq:void} is zero according to Eq. \ref{ec:campoEthetav}, which can be taken also as a good approximation to the very small values obtained with the FEM calculation.

It must be noted that the boundary values shown in figure \ref{fig:conds_contorno}b 
 have also an experimental counterpart: the inductance of the driver in the absence of plasma, $L_\mathrm{d}$. The FEM calculation gives a driver inductance $L_\mathrm{d}^\mathrm{FEM} = 9.4$ $\mu$H, in quite good agreement with the measured $L_\mathrm{d}=9.6$ $\mu$H \cite{jain2022investigation-o}. This value cannot be attained with the 2D electromagnetic code because of the limited calculation domain. It is possible, however, to define a reduced value using the flux linked by the RF coils considering only the plasma region. In vacuum we obtain $L_{\downarrow} = 6.2$ $\mu$H. Since, according to figure \ref{fig:conds_contorno}b, this value of $L_{\downarrow}$ must be practically the same using the FEM calculation, we can use it as an estimate of the inductance compatible with $L_\mathrm{d}^\mathrm{FEM}$ and, consequently, with $L_\mathrm{d}$. In what follows we shall refer to this reduced value $L_\downarrow$.

For compatibility with the boundary condition at the cylindrical surface of the Faraday shield, we must have $E_\theta^\mathrm{void} (r_\mathrm{p},z_\mathrm{end})=0$ at this edge of the opening to the expansion region. A possibility is to take a null field all along this region of the boundary, $E_\theta^\mathrm{void} (r,z_\mathrm{end})=0$. There are two arguments in favor of this choice. First, according to the FEM calculation, the void-driver field at this boundary is comparatively small. Second, not only the theoretical vacuum field $E_\theta^\mathrm{V}(r,z_\mathrm{end})$ is also considerably reduced in this boundary region, but the iterative process described below tends to make it evanesce. This has been verified considering only one step function in $z=z_\mathrm{ini}$, that is, $s(z)=\tanh [(z-z_\mathrm{ini})/2\delta]$, thus leaving the other end with the value $E_\theta^\mathrm{void}(r,z_\mathrm{end})=f_\mathrm{c}E_\theta^\mathrm{V}$, as also indicated in figure \ref{fig:conds_contorno}. Since, as mentioned, this option renders $E_\theta^\mathrm{void}(r,z_\mathrm{end})$ quite small,  and this boundary field decreases even further in presence of the plasma, we have simplified the calculations that follow with the fixed boundary condition $E_\theta^\mathrm{void}(r,z_\mathrm{end})=0$. In summary, the boundary conditions applied to the void driver are
\begin{eqnarray}
E_\theta^\mathrm{void}(0,z) &=& 0\label{eq:bndry1}\\
E_\theta^\mathrm{void}(r,0) &=& 0\label{eq:bndry2}\\
E_\theta^\mathrm{void}(r_\mathrm{p},z) &=& f_\mathrm{c}s(z)E_\theta^\mathrm{V}(r_\mathrm{p},z) \label{eq:bndry3}\\
E_\theta^\mathrm{void}(r,z_\mathrm{end}) &=& 0,\label{eq:bndry4}
\end{eqnarray}
with $s(z)$ as in Eq.~\ref{eq:shaping function} and $f_\mathrm{c}=0.38$ to match the void driver inductance.

\paragraph{Iterations with plasma}

Once the boundary conditions for the void driver have been set, we can use them as initial guess when there is a plasma that reacts with  induced currents. Here it is very important to keep in mind that the experimental inductance of the drivers decreases very little in presence of the plasma (Sec.~\ref{subsec:electrical}).
It could be argued that, in consequence, an iterative process that refines the boundary conditions according to the plasma response is barely necessary. This would be true if we knew a precise formula for the plasma conductivity, which is not the case. On the contrary, we seek a description of the plasma that, based on experimental profiles, gives a plasma reaction that is compatible with the experimental knowledge of RF power delivered and approximately constant inductance. Such is the purpose of the iterative process.
 The essential information that will be gained from the calculations consists of the absorbed power and the decrement of the inductance (directly related with the value of the electric field at $r=r_\mathrm{p}$) in presence of the plasma. 

The iterative process begins with the boundary conditions Eq.~\ref{eq:bndry1}--\ref{eq:bndry4}. The driver inductance nears the void-driver value $L_\mathrm{d}$, to which we have associated a calculated $L_{\downarrow}$ considering only the plasma region. Successive steps in the iterative process are as follows:
 \begin{enumerate}
\item The initial $E_\theta (r,z)$ provides a distribution of plasma current densities $J_\theta(r,z) = \sigma(r,z) E_\theta (r,z)$ that, given its symmetry, can be considered as a set of current-carrying loops located at positions $(R_{\mathrm{b}_i},z_{\mathrm{b}_i})$ for each $i$-th loop, from which a contribution Eq.~\ref{ec:campoEthetav} can be calculated at the boundary points $(r_{\mathrm{p}},z)$ and, eventually, $(r,z_{\mathrm{end}})$.
\item The new boundary values consist of the initial boundary field corrected by the contribution from the plasma current loops.
\item The process is repeated until the correction to the boundary $E_\theta$ is considered negligible.
\end{enumerate}

Let us change slightly the notation to better describe the iterative process. We use the symbol $\partial$ for the boundary  of the calculation domain, where the electric field (we recall it only has azimuthal component), $E_\partial$, depends on the currents (electric fields) in the whole plasma, $E(\mathbf{x})$. We symbolize this by writing $E_\partial[E(\mathbf{x})]$. The process starts with a value $E_\partial^\mathrm{void}$. The field $E(\mathbf{x})$ is calculated in consecutive iterations: the $j$-th one yields $E^j(\mathbf{x})$ and a corresponding $E_\partial^{j+1} =E_\partial^\mathrm{void}+\Delta E_\partial^j$, where 
 $\Delta E_\partial^j[E^{j}(\mathbf{x})]$ is the contribution of the calculated currents to the boundary that will be used in the next iterative step. Note that it does not make sense calculating  $E_\partial^{j+1} = E_\partial^{j} + \Delta E_\partial^j[E^{j}(\mathbf{x})]$ because, in such case, the convergence is obtained when $E_\partial^{j+1} \approx E_\partial^{j}$, which implies $\Delta E_\partial^j[E^{j}(\mathbf{x})] \to 0$ as the iterations grow. A field that does not modify the boundary values through the associated currents is null unless the conductivity is negligible. Therefore, we impose
\begin{equation}
E_\partial^{j+1} = E_\partial^\mathrm{void} + \Delta E_\partial^j[E^{j}(\mathbf{x})],
\label{ec:FEC2018_14_1}
\end{equation}
so the electric field at the boundary is updated by adding the plasma response to the fixed vacuum field. This process converges when the field at the boundary is such that $E_\partial  \approx E_\partial^\mathrm{void} + \Delta E_\partial$. Observe that $\Delta E_\partial$ gives a ``negative'' contribution as it opposes the vacuum field. Therefore, the condition can be expressed by saying that the converged field is the vacuum one minus the effect of the currents provoked by the converged field itself.

Since the power is a volume integral of $|E(\mathbf{x})|^2$, the convergence of the electric field provides a convergence of the ohmic power towards a value that can be used to define a criterion for convergence. When the relative change of ohmic power is below some value, say 1\%, the iterative process stops.  The convergence is not guaranteed if the conductivity values are too high because an excessive response of the plasma, i.e., too high induced currents in the first step, could overcome the boundary values giving rise to an unphysical amplification of the currents. With acceptable values of the conductivity, the iterations converge in a few steps. In such case, the electric field in the open side is reduced to negligible values if it is let to evolve as indicated in the previous paragraph. For this reason, it is justified using also a null value of the electric field in the expansion region side of the boundary.

\subsubsection{Currents, power and inductance}

As before, we drop the sub-index $\theta$ for the only component of both, electric field and current density. The latter in this problem has a harmonic character, $J(r,z;t) = \sigma E(r,z) \exp(\imath \omega t)$ with a complex conductivity $\sigma = |\sigma| \exp (\imath \phi )$. Thus, we have a current density
\begin{equation}
J(r,z;t) = |\sigma (r,z)|E(r,z)e^{\imath [\omega t + \phi (r,z)]}
\label{ec:lido64_3}
\end{equation}
with a complex field amplitude $E(r,z)=|E(r,z)|\exp (\imath \varphi)$. Consequently, the current density
\begin{equation}
J(r,z;t) = |\sigma (r,z)| |E(r,z)| e^{\imath [ \varphi(r,z) + \phi (r,z)]}e^{\imath \omega t}
\label{ec:lido64_3b}
\end{equation}
 is oscillating with a phase $\imath [\omega t + \phi (r,z)+\varphi(r,z)]$ and an amplitude $|J| = |\sigma (r,z)| |E(r,z)|
$. Each current loop gives rise to a contribution to the boundary
\begin{equation}
E_{\partial,I(r,z)}  \equiv -\imath f G(k) I(r,z)  e^{\imath [ \varphi(r,z) + \phi (r,z)]}e^{\imath \omega t},
\label{ec:lido67_5}
\end{equation}
where Eq.~\ref{ec:lido64_3b} indicates that the current in each loop of infinitesimal cross section $\dd r \dd z$, including its spatial phase, is
\begin{equation}
\dd I(r,z) = |\sigma (r,z)| |E(r,z)| e^{\imath [ \varphi(r,z) + \phi (r,z)]} \dd r \dd z.
\label{ec:lido67_5b}
\end{equation}
These currents can be used to obtain the ohmic power dissipated by the plasma due to the induced field. The one-period average for a given circuit under harmonic drive is
\begin{equation}
\langle P \rangle = \frac{1}{2}\Re\{I^*V\},
\label{ec:lido71_2}
\end{equation}
where $V$ represents the induced voltage related with the circulation of the electric field, $I^*$ is the complex conjugate of the current and $\Re$ takes the real part of its argument. For each loop of plasma current, owing to the symmetry of the problem, we have a complex voltage $V(r,z) = -2\pi r E (r,z)$. 
The complex conjugate of Eq.~\ref{ec:lido67_5b} can be used for the current.

The self-induction coefficient is, by definition, $L = \dd \Phi/\dd I_\mathrm{RF}$, 
where the magnetic flux is the circulation of the magnetic vector potential and $I_\mathrm{RF}$ is the current in the external winding. The flux through one circular loop of radius $r$ is
\begin{equation}
\Phi = \oint A \dd l =2\pi r A = \imath \frac{r}{f}E,
\label{ec:lido72_3}
\end{equation}
where again we take advantage of the symmetry of the problem and we have substituted $A = ( \imath/\omega){E}$.

We have modified the theoretical vacuum field in order to have a good approximation for the void driver, but restricted to the calculation domain, up to $r=r_\mathrm{p} < R_\mathrm{b}$. Consequently, we can only provide the reduced inductance, $L_\downarrow$.  
 Then, our approximation to the inductance for a set of $N_\mathrm{b}$ loops centered at $z_{\mathrm{b}i}$ is
\begin{equation}
L_\downarrow = \sum_{i=1}^{N_\mathrm{b}} \frac{r_\mathrm{p}}{I_\mathrm{RF}f} \Re \{ \imath E (r_\mathrm{p},z_{\mathrm{b}i})\},
\label{ec:lido72_5}
\end{equation}
which includes the plasma effects after the iteration process. Here we remind that, even if $L_\downarrow$ is an under-estimate of the inductance, we only need to know the modification of this value in comparison with equivalent void-driver inductance, $L^\mathrm{void}_\downarrow$, obtained when $E^\mathrm{void}$ is substituted in Eq. \ref{ec:lido72_5}.


\section{\label{sec:datos experimentales}Experimental data}

\subsection{Electron density and temperature}

For the calculations that follow we take data from the S16 campaign of the SPIDER device, which was devoted to characterize the plasmas inside the drivers and the adjacent expansion region. Most of the information that we use below was obtained by inserting probes in these regions \cite{sartori2021development-of-}. Different probes, often located in different drivers, have fixed radial locations (our $r$-coordinate). Movable probes along the axial direction were used to scan plasma parameters in our $z$-coordinate, see figure \ref{fig:conds_contorno}. The parameter space for these experiments is very large, including external knobs like power, gas pressure, bias voltage, filter magnetic field intensity etc.  Therefore, due to the natural limitations of the experimental campaign, the available data are also limited in terms of complete $(r,z)$ maps of electron density and temperature.

\begin{figure}[htbp] 
   \centering
   \includegraphics[width=0.95\columnwidth]{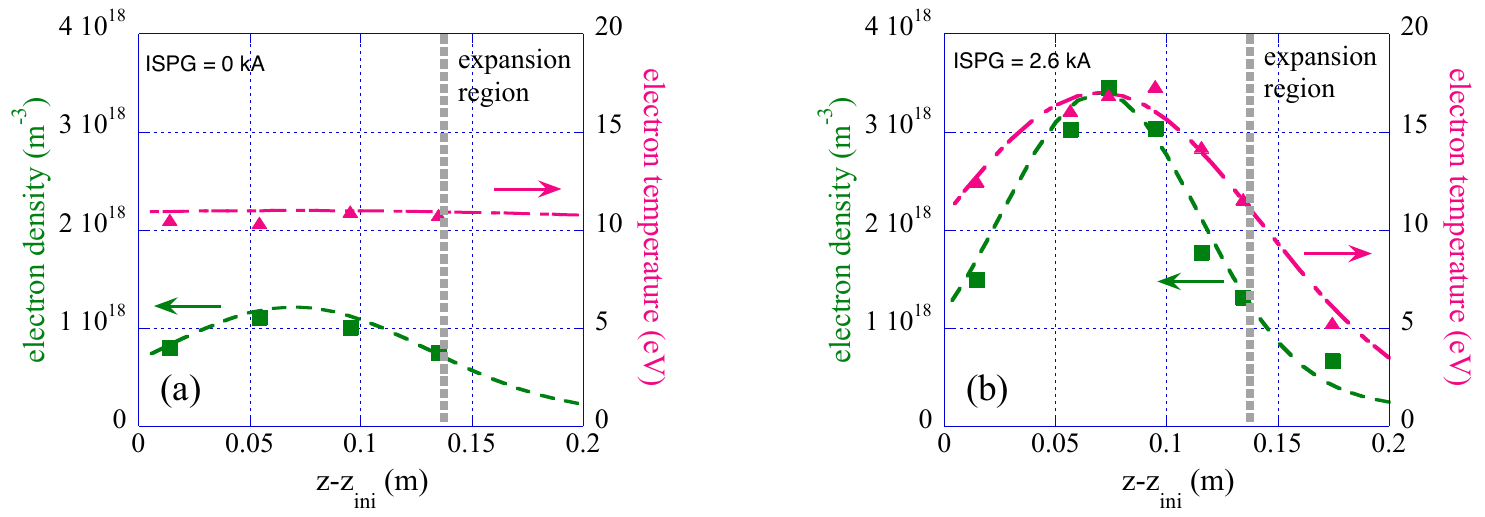} 
   \caption{ Probe data for the electron density (squares) and temperature (tringles) taken along the driver axis from SPIDER discharges without (a) and with (b) filter magnetic field.}
   \label{fig:olla68_1}
\end{figure}

 We describe with Gaussian functions the electron density and temperature profiles obtained from measurements in two operating conditions: without ($I_\mathrm{PG}=0$) and with ($I_\mathrm{PG}=2.6$ kA; $B_\mathrm{f} \approx 4$ mT) filter magnetic field \cite{toigo2021on-the-road-to-}, where $I_\mathrm{PG}$ is the current in the plasma grid circuit providing the filter field, $B_\mathrm{f}$.
  Figure \ref{fig:olla68_1} shows, with symbols, the electron densities and temperatures obtained along the driver axis as a function of the distance from the back side of the driver (see figure \ref{fig:conds_contorno}a). The dashed lines are the functions that will be used in the calculations. The variation along the $z$-axis is acceptably well defined for both types of discharge. The plasma profiles for the $I_\mathrm{PG}=2.6$ kA case have also some data along the radial direction (see figure \ref{fig:profiles_ISPG}, to be discussed later).  Thus, if $\mathcal{G}_y(\Delta_y,y_c) \equiv \exp [(y-y_\mathrm{c})/2\Delta_y^2]$ represents a  $\Delta_y$-width, $y_c$-centered Gaussian function on the variable $y$, we  represent the density and temperature spatial distributions of this case with the 2D functions
 \begin{equation}
\left .
\begin{array}{c}
     n_e(r,z)=[3.2\mathcal{G}_r(0.05,0)\mathcal{G}_z(0.1,z_\mathrm{ini}+0.07) + 0.2]\times 10^{18} \mbox{ m}^{-3}\\
     T_e(r,z)=17 \mathcal{G}_r(0.103,0)\mathcal{G}_z(0.1,z_\mathrm{ini}+0.07) \mbox{ eV},
\end{array}
\right \}
\label{ec:lido76_2}
\end{equation}
where the centering of $\mathcal{G}_z$ imposes the peaking near the center of the driver in the axial dimension. All distances are given in meters.

\begin{figure}[htbp] 
   \centering
   \includegraphics[width=0.37\columnwidth]{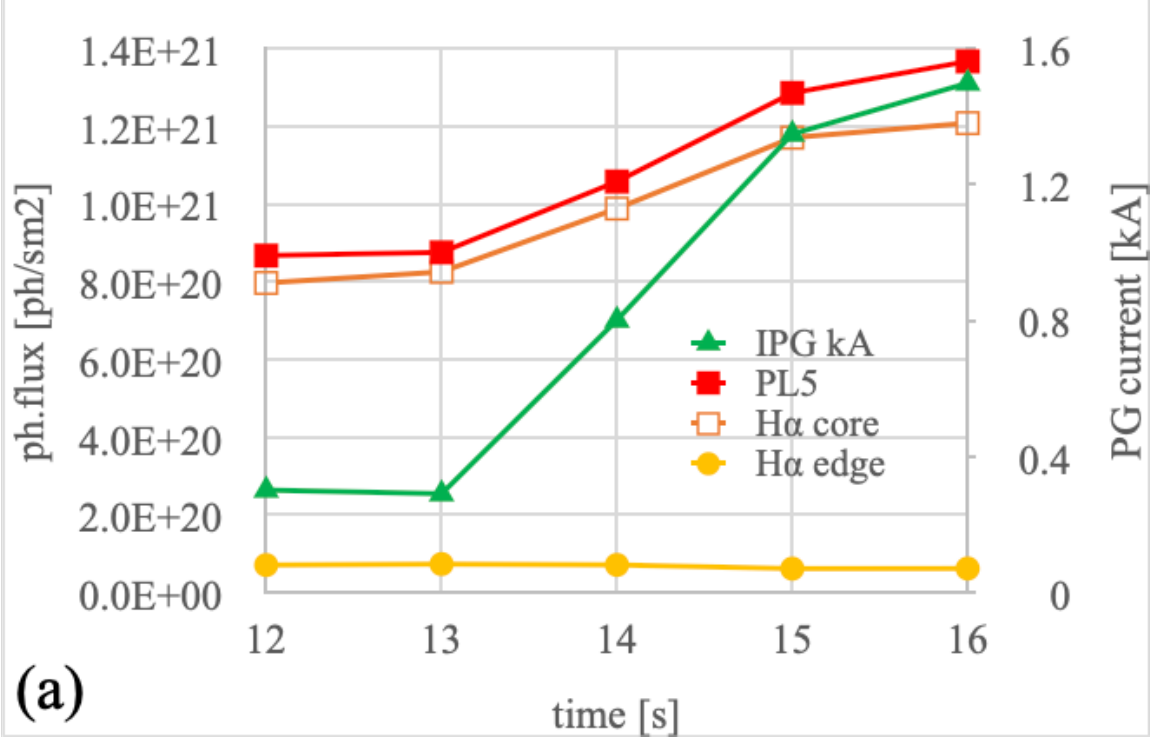} 
   \includegraphics[width=0.29\columnwidth]{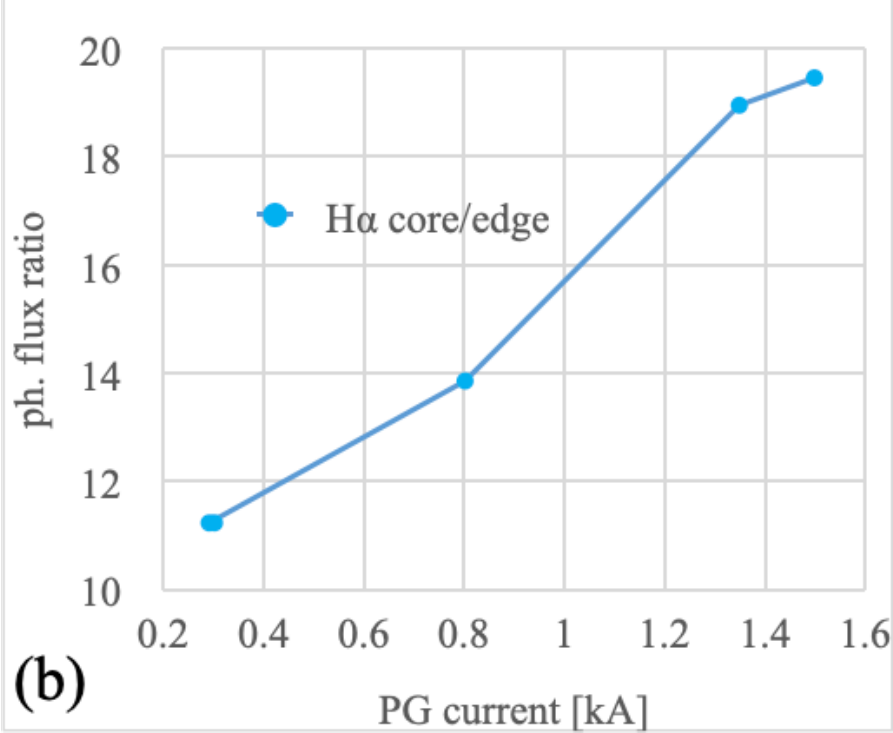} 
   \includegraphics[width=0.29\columnwidth]{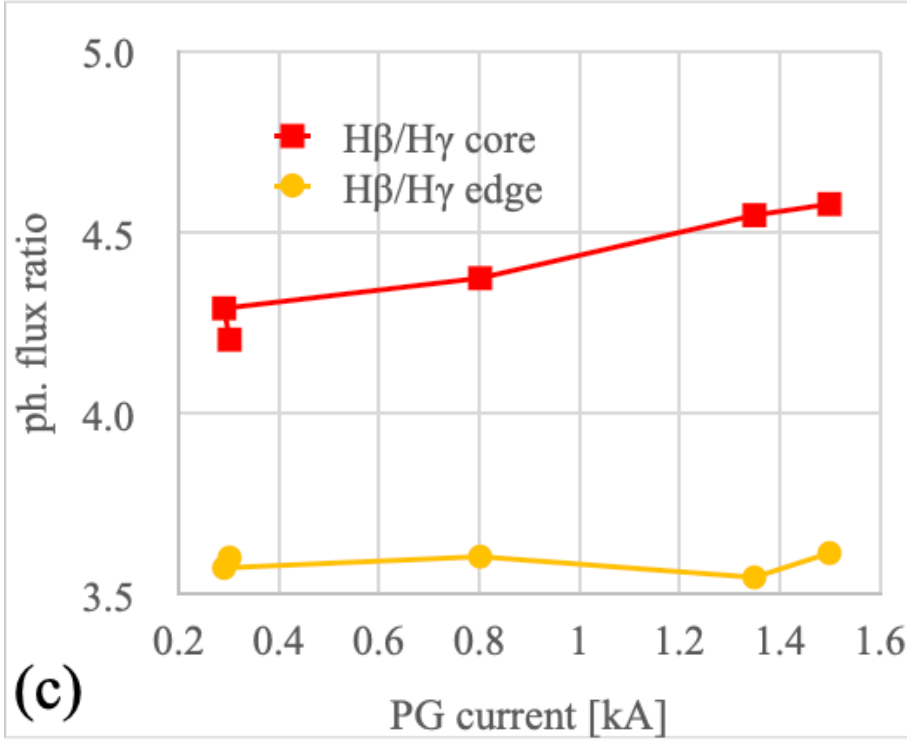} 
   \caption{ Spectroscopy data from a SPIDER discharge with varying plasma grid current ($I_\mathrm{PG}$), which controls the intensity of the filter magnetic field. (a) Time evolution of $I_\mathrm{PG}$ (triangles) and H$_\alpha$ signals collected from the core (squares, see text) and radial edge (dots) plasma regions during the $I_\mathrm{PG}$ ramp. (b) Ratio between core and edge intensities of H$_\alpha$ light as $I_\mathrm{PG}$ is changed. (c) Ratio between the $\beta$ and $\gamma$ lines --Balmer ratio-- in core (squares) and edge (dots) chords.}
   \label{fig:spectroscopy}
\end{figure}

\begin{figure}[htbp] 
   \centering
   \includegraphics[width=0.85\columnwidth]{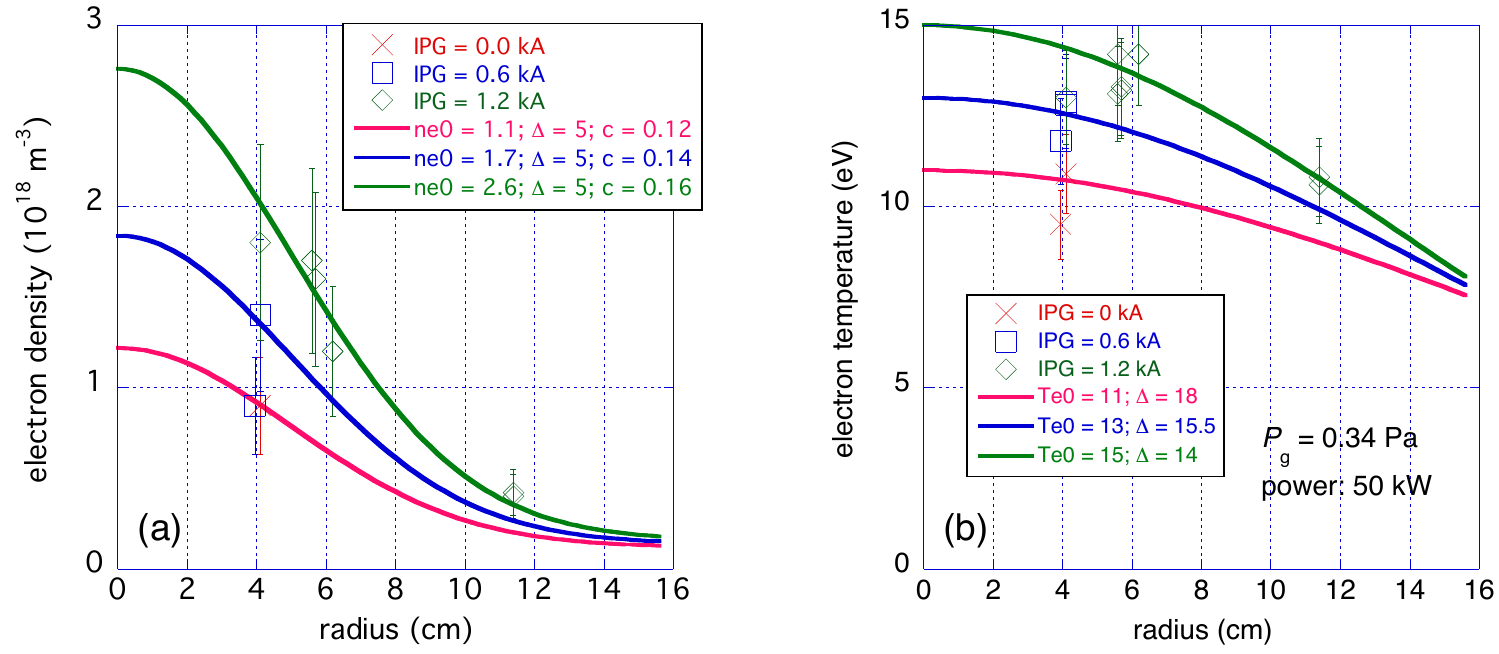}
   \caption{Probe data of the electron density (a) and temperature (b) at different radial locations near the center of the driver ($z=0$) for different PG currents (symbols).  The corresponding Gaussian functions used to represent the profiles are shown with lines. Their value at $r=0$ ($10^{18}$ m$^{-3}$, eV), width $\Delta$ (cm) and possible offset $c$  ($10^{18}$ m$^{-3}$) are indicated in the labels.}
   \label{fig:ISPGbajoB}
\end{figure}

Unfortunately, the $I_\mathrm{PG}=0$ case does not have information on the radial profiles, but only along the $z$-coordinate. This case is, however, important precisely because there is no contribution from the static filter magnetic field, thus becoming the natural control case to study the effects of this field on the plasma conductivity. Here we have used supplementary information from other discharges at low $I_\mathrm{PG}$ and from spectroscopic measurements. Figure \ref{fig:spectroscopy} (a) shows
 the evolution of $I_\mathrm{PG}$ during the time of the discharge, along with the variation of H$_\alpha$ in core and edge chords. The two core lines correspond to different measurements, from a photodiode that collects H$_\alpha$ light after the corresponding filter (empty squares) and from a spectrometer (filled squares), and are shown to give an indication of the consistency of the measurements.
 According to this figure, the H$_\alpha$ light emission increases with the $I_\mathrm{PG}$ current (intensity of the filter magnetic field) in the core of the driver, but not in the edge region ($r \approx 0.7r_\mathrm{p}$), thus increasing the peaking factor (b) in agreement with the peaking of the electron density measured with probes. At the same time, the Balmer ratio (c) of the edge emission stays approximately constant during the $I_\mathrm{PG}$ scan. Simultaneous constant edge H$_\alpha$ light and Balmer ratio are hard to explain if the density and temperature of the edge region are not approximately constant during the $I_\mathrm{PG}$ scan. 
 Therefore, both electron density and temperature are assumed to remain approximately unchanged (other parameters fixed, as is our case) in the outer region as the filter field changes.  Figure \ref{fig:ISPGbajoB} shows experimental data from plasmas at the lowest $I_\mathrm{PG}$, together with the Gaussian functions used to represent the radial profiles. 
 The error-bars in the figure were set at 30\% on the density, mostly given by the uncertainty on the effective mass of the positive ions (between 1 and 2 amu in the driver); and 10\% on the temperature, to give an estimate of the data dispersion.
 We note that only the $I_\mathrm{PG} = 1.2$ kA case has data near the edge radial region. The function adopted for the $I_\mathrm{PG}= 0$ kA case has been obtained for comparison with the other two cases shown and using the above qualitative information from spectroscopy. In consequence, now we use
\begin{equation}
\left .
\begin{array}{c}
     n_e(r,z)=[1.1\mathcal{G}_r(0.05,0)\mathcal{G}_z(0.09,z_\mathrm{ini}+0.07) + 0.12]\times 10^{18} \mbox{ m}^{-3}\\
     T_e(r,z)=11 \mathcal{G}_r(0.18,0)\mathcal{G}_z(0.60,z_\mathrm{ini}+0.07) \mbox{ eV},
\end{array}
\right \}
\label{ec:lido76_1}
\end{equation}
which are the functions represented for the $I_\mathrm{PG}=0$ case in figure \ref{fig:ISPGbajoB}. 
Note that an almost flat $T_e$ is taken in the $z$ dimension, in agreement with the experiments (figure \ref{fig:olla68_1}). This is a notable change between discharges without and with filter magnetic field: the latter concentrate the hot electrons near the center of the driver region.
 Other values are common: $P_\mathrm{g}=0.34$ Pa of neutral gas pressure, $1$ MHz for the RF and a one-driver nominal power $P_\mathrm{RF}=50$ kW.

\subsection{\label{subsec:electrical}Electrical parameters}

The numerical problem to be solved is based on the knowledge of two electrical parameters: the input power and the inductance. 
The measured impedance in the drivers of the SPIDER device without plasma is considered to have negligible capacitive reactance, $Z = R_\mathrm{d} + i \omega L_\mathrm{d}$ with resistive part $R_\mathrm{d} \sim 2$ $\Omega$ and inductive part $L_\mathrm{d} \sim 10$ $\mu$H \cite{Recchia2021studies-on-powe}. We have seen (Sec.~\ref{ecs y cc contorno}) that $L_\mathrm{d}$ is used to define initial boundary conditions in solving Eq.~\ref{ec:ecuacion em}. Recent studies \cite{jain2022investigation-o} with improved electrical diagnostics indicate that a representative value of the vacuum impedance of the drivers in SPIDER is $L_\mathrm{d} \approx 9.6$ $\mu$H.  In addition, and in agreement with previous results, the variations of the driver inductance have been found on the order of 1\% in different operating conditions of gas pressure, RF power, plasma-grid current etc. 
 This small variation is the main information for our purposes.

Given the fixed RF frequency, the only electrical input in the calculations is the amplitude of the current in the RF winding, $I_\mathrm{RF}$. This value cannot be taken directly from the electrical measurements at the output of the RF generators. Aside from the RF power, $I_\mathrm{RF}$ is a consequence of the estimate of plasma equivalent resistance and depends, consequently, on the plasma response. 
 Recent studies \cite{jain2022investigation-o} provide this estimate for shots at nominal power 45 kW and gas pressure $P_\mathrm{g}= 0.4$ Pa, similar to our conditions, in a range of plasma-grid currents from $I_\mathrm{PG}=1$ to $I_\mathrm{PG}=2.5$ kA. 
  Based on this reference, and recalling the two types of plasma described in Sec.~\ref{sec:datos experimentales}, we have fixed $I_\mathrm{RF}/\sqrt{2}=\sqrt{P_\mathrm{RF}/R_\mathrm{d}}$ using $R_\mathrm{d}=2.1$ $\Omega$ for our case with magnetic filter field, $I_\mathrm{PG}=2.6$ kA. The case without filter magnetic field is obtained by extrapolation of the trend found, $R_\mathrm{d}=1.7$ $\Omega$. Therefore, we set the following values: $I_\mathrm{PG} = 0$ A (no filter field), $I_\mathrm{RF}=242$ kA; and $I_\mathrm{PG} = 2.6$ A (with filter field), $I_\mathrm{RF}=218$ A. 

Note that, since we are fixing $I_\mathrm{RF}$ according to the estimates of the plasma equivalent resistance for a given input power (50 kW), this work is different to the reference \cite{Recchia2021studies-on-powe} in the sense that $I_\mathrm{RF}$ is an input and the absorbed power is an output. In other words, the present 2D electromagnetic calculations provide values of the driver efficiency taking $I_\mathrm{RF}$ and the net power as inputs.

\section{\label{sec:conductividad}Plasma conductivity}
 

A convenient way to obtain the conductivity of a medium characterized by fluid equations consists of writing the linear momentum balance for the electron fluid and solving for the velocity. Here it is assumed that the currents are due to the reaction of the electron fluid alone. The definition of the current density allows then obtaining an expression for the conductivity. This can be easily done in planar geometry for a harmonic drive when advection, diamagnetic forces and magnetic field effects are neglected. Considering a drag force as proportional to the electron fluid velocity via a collision frequency $\nu$, one obtains the well known formula (see, for instance, \cite{Chabert2011Physics-of-Radi}),
\begin{equation}
\sigma = \frac{n_e e^2}{m_e( \nu + \imath \omega)},
\label{ec:conductividad_aprox}
\end{equation}
where $m_e$, $e$ and $n_e$ are, respectively, mass, charge magnitude and density of  the electron species. The generic meaning of $\nu$ as quantifier of the drag allows to consider an effective collisionality that takes into account processes of possibly different nature. In \cite{Jain2018Evaluation-of-p,Jain2018Improved-Method}, cross-section data for the local collisional processes between electrons and neutrals were used to obtain reaction rates and then collisionalities for the elastic, $\nu_{en}^p$, and inelastic by ionization, $\nu_{en}^\mathrm{i}$, processes. With the addition of the electron-ion elastic collisions, $\nu_{ei}$, we would have $\nu = \nu_{en}^p+\nu_{en}^\mathrm{i}+\nu_{ei}$ in formula \ref{ec:conductividad_aprox}. In the present work we have taken the total cross-section for collisions (elastic and inelastic) between the electrons and Hydrogen neutrals published in \cite{Yoon2008Cross-Sections-}. The reaction rate that corresponds to this cross-section, $\sigma_{en}^\mathrm{tot}$, is obtained considering, as usual, a Maxwellian distribution of energies,
\begin{equation}
k^\mathrm{tot}_{en} = \sqrt{\frac{8}{\pi m_eT_e^3}} \int_0^\infty \mathcal{E} \sigma_{en}^\mathrm{tot} e^{\mathcal{E}/T_e} \dd \mathcal{E}.
\label{ec:reaction rates}
\end{equation}
With the values thus obtained in a range of temperatures (energies), $0.01 \leq T_e \leq 100$ eV, we have obtained the fit that will be used for convenience in the calculation of the local collisionalities involving electron-neutral collisions. Adjusting the reaction rates to a fourth order polynomial,
\begin{equation}
\log k_{en}^\mathrm{tot} = \sum_{j=0}^4 c_i (\log T_e)^j,
\label{ec:polynomial fit}
\end{equation}
 the set of coefficients is $(c_0,c_1,c_2,c_3,c_4) = (-12.9736, 0.5714, -0.4728, 0.1419, -0.0153)$. 

\begin{figure}[htp] 
   \centering
   \includegraphics[width=0.6\columnwidth]{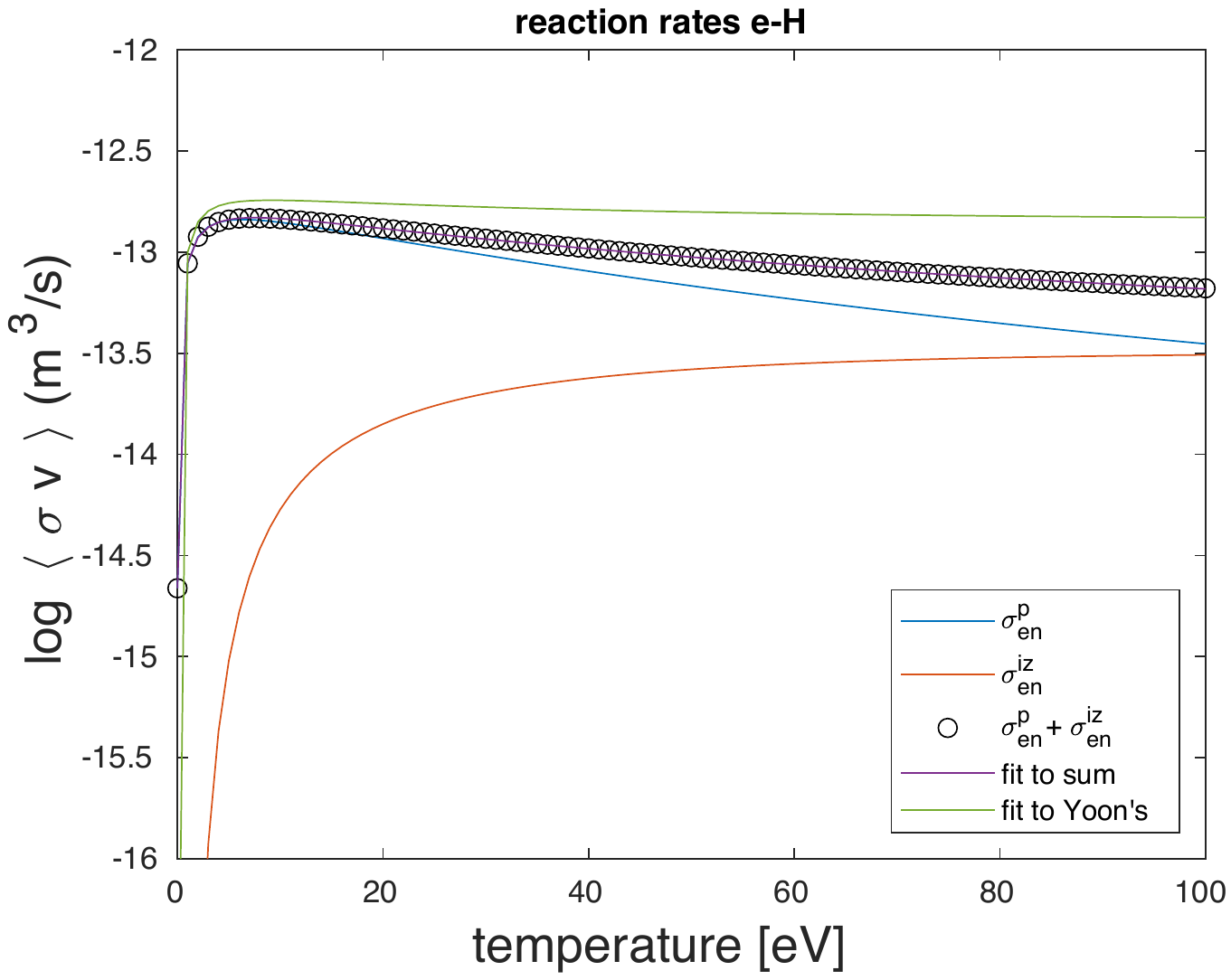} 
   \caption{Reaction rates obtained from the cross-sections associated to several collisional processes of electrons in a Hydrogen gas: elastic collisions, $\sigma_{en}^p$; ionizing collisions, $\sigma_{en}^\mathrm{i}$; their sum (symbols) along with its polynomial fit Eq.~\ref{ec:polynomial fit}; and a similar fit to the recommended total cross-sections published by Yoon et al.~\cite{Yoon2008Cross-Sections-}.}
   \label{fig:secciones_ef_col_el}
\end{figure}

The total local collisionality between electrons and neutrals, $\nu_{en}^\mathrm{tot} = n_\mathrm{g} k_{en}^\mathrm{tot}$ where $n_\mathrm{g}$ is the neutral gas density, is added to the known electron-ion collision frequency to obtain a total local collisionality
\begin{equation}
\nu_\mathrm{local}= \nu_{en}^\mathrm{tot} + \nu_{ei}.
\label{ec:nu local}
\end{equation}
Figure \ref{fig:secciones_ef_col_el} shows the reaction rates Eq.~\ref{ec:reaction rates} for the electron-neutral cross-sections 
 $\nu_{en}^p$ and $\nu_{en}^\mathrm{i}$. Their sum and its 4th order polynomial fit show that Eq.~\ref{ec:polynomial fit} correctly accounts for the set of numerical integrals. For comparison, we also show our chosen fit to the total electron-neutral cross-sections provided in Ref.  
 \cite{Yoon2008Cross-Sections-}, see coefficients above, which includes many more collisional processes. As can be seen, reducing the collisions between electrons and Hydrogen neutrals to the sum $\nu_{en}^p+\nu_{en}^\mathrm{i}$ gives a rather good approximation in the range of temperatures of interest to these studies, $\sim 10$ eV.

\begin{figure}[htp] 
   \centering
   \includegraphics[width=0.42\columnwidth]{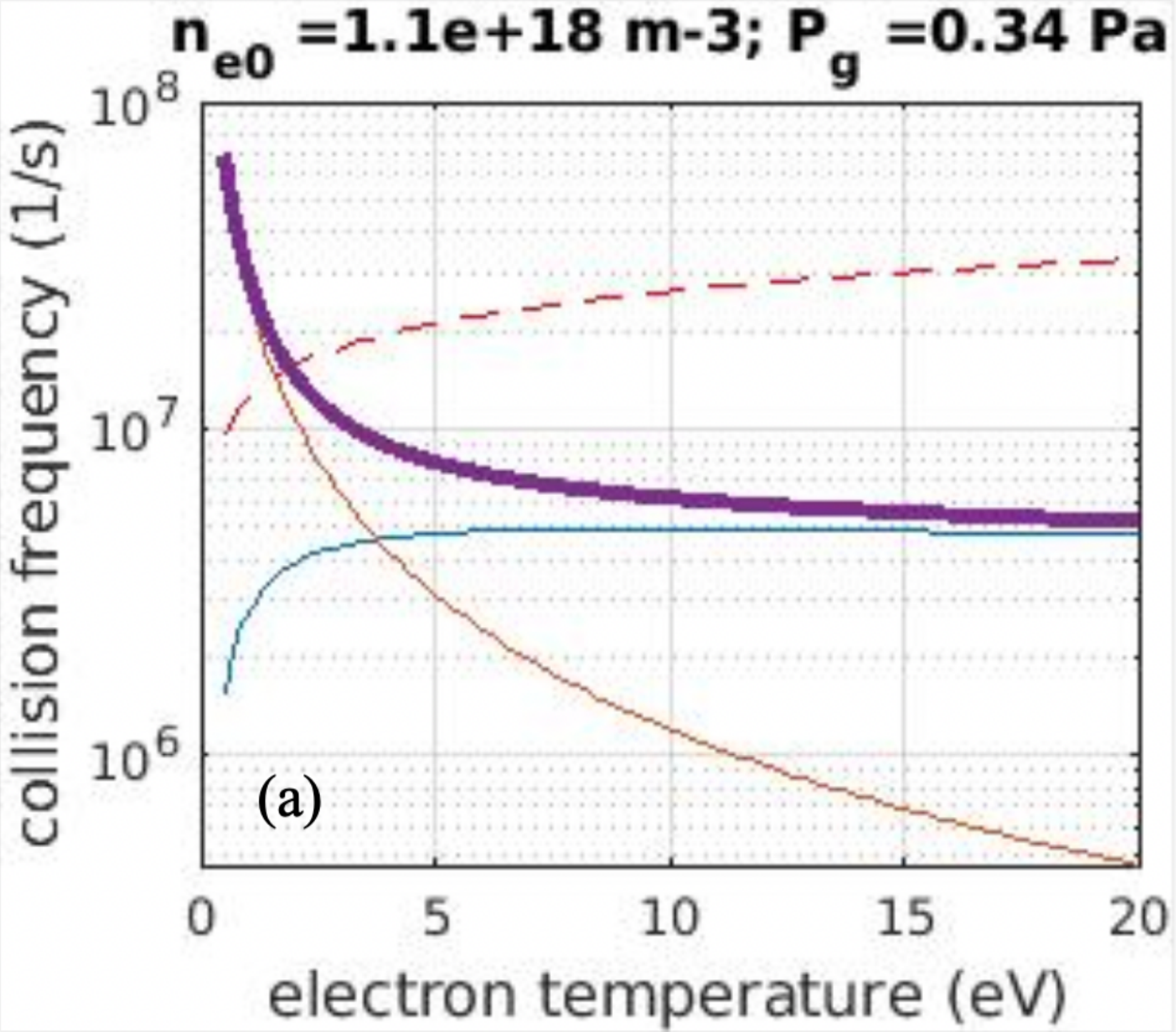} 
   \includegraphics[width=0.47\columnwidth]{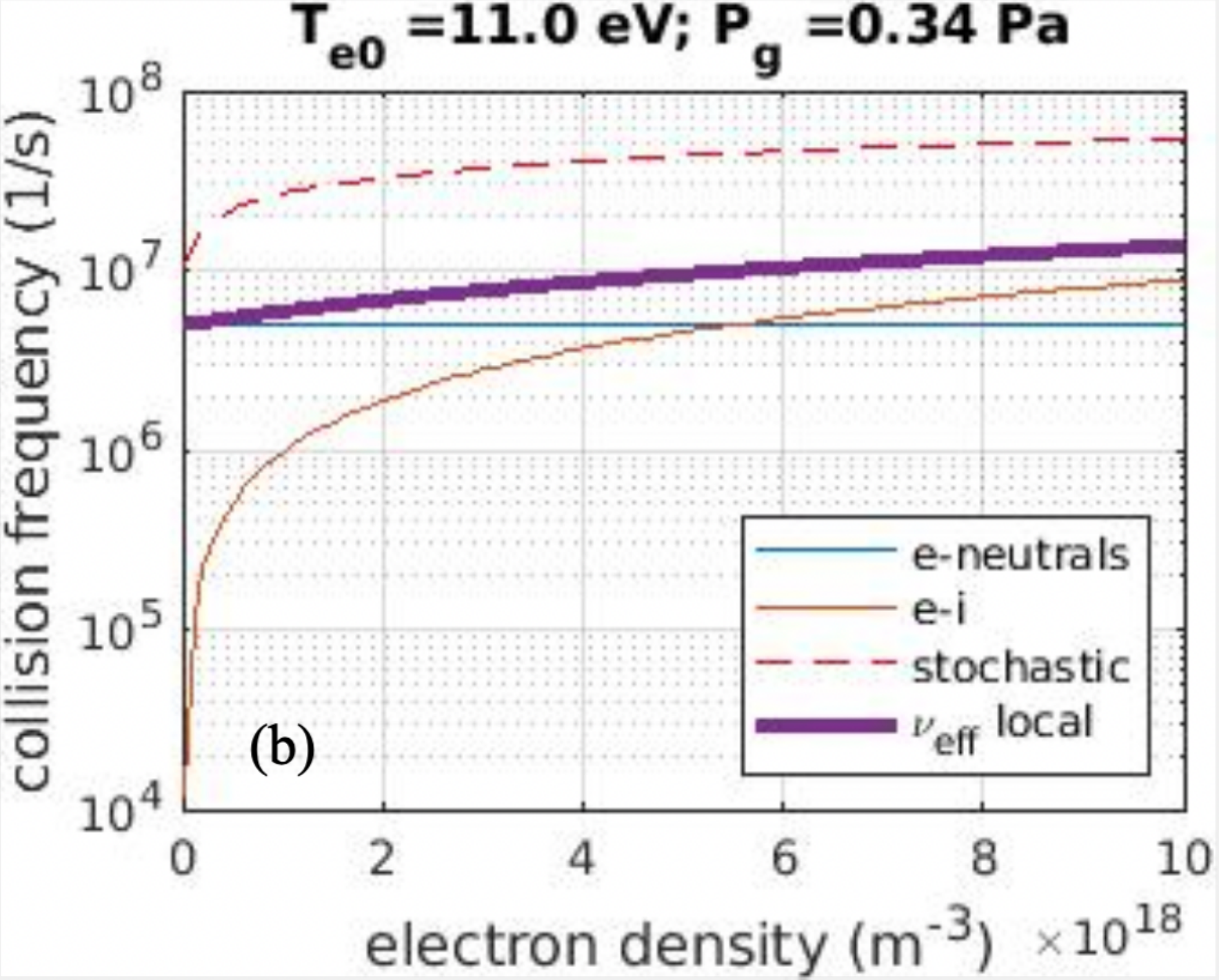} 
   \caption{Collision frequencies obtained for a Hydrogen gas pressure $P_\mathrm{g}=0.34$ Pa as a function of the electron temperature (a) and density (b). The local collisionalities (thin lines) are added to obtain an effective collisionality (thick lines) to be compared with the collisionality associated to stochastic heating (dashed lines).}
   \label{fig:barridos_nu}
\end{figure}

 Many works have indicated that collisionless heating can become, depending on the plasma conditions, the main heating mechanism in inductively coupled plasmas.  This type of heating, often called ``stochastic heating'', is produced when the electrons explore zones of considerably different electric field during one RF period. This can only happen if, in turn, other collisional processes have low enough frequencies, as it can easily happen at low neutral gas pressures. In terms of conductivity, this kind of heating can be associated to an equivalent collisionality, $\nu_\mathrm{st}$, which is obtained by equating the collisionless heating power to an effective collisional heating power. The calculations are involved and, strictly speaking, should consider the non-locality by evaluating the surroundings of each point. In this way, the corresponding conductivity should be expressed via integral expressions (e.g. \cite{kaganovich2003self-consistent,turner2009collisionless-h}). 
As a first approximation to the problem, here we adopt the formulation based on the same expressions for the skin depth and equivalent collisionality used in \cite{cazzador2015semi-analytical,Jain2018Evaluation-of-p,Jain2018Improved-Method,Recchia2021studies-on-powe}. These are based on the limits of $\nu_\mathrm{st}/\omega$ developed in \cite{Vahedi1995Analytic-model-} to obtain $\nu_\mathrm{st}$ as a function of the skin depth, $\delta$, and the ratio $\alpha$ between the transit time of a thermal electron throught the skin depth and the RF period. The parameter $\alpha = 4\delta^2\omega^2/\pi v_\mathrm{th}^2 \ll 1$ always in our case, where $v_\mathrm{th}$ is the thermal speed. Therefore we iterate the collisionality starting with $\nu_0 = \nu_\mathrm{eff}^\mathrm{local}=\nu_{en}^\mathrm{tot} + \nu_{ei}$ to obtain $\delta_0 \equiv\delta(\nu_0)$, from here $\alpha_0 \equiv \alpha(\delta_0)$ and then the successive $k$-th iteration values
$$
\nu_k = \frac{1}{2\pi}\frac{v_\mathrm{th}}{\delta_{k-1}}.
$$
The process converges quite rapidly, in some five iterations depending on parameters. 
 The resulting collisionalities are shown in figure \ref{fig:barridos_nu} with dashed lines and compared with the local values. It can be clearly appreciated that the collision frequencies associated to stochastic heating are dominant except at the lowest electron temperatures. The figures have been obtained using the gas pressure of the SPIDER experiments from which we have obtained the plasma profiles (Sec. \ref{sec:datos experimentales}). It should be kept in mind that these evaluations are considered as having the appropriate order of magnitude. More detailed calculations might be necessary for an eventual comparison with kinetic codes or with detailed experimental data.

It must be warned that a main-ion density close to the electron density gives ion plasma frequencies quite comparable to the RF frequency in SPIDER discharges; and even higher, depending on plasma parameters. In consequence it should be kept in mind that this work concentrates on the very relevant electron dynamics, which gives rise to the conductivity models exposed above. Further refinements might require considering ion dynamics as well.


\section{\label{sec:resultados} Calculations}

Eq.~\ref{ec:ecuacion em} is written in cylindrical coordinates (Appendix \ref{apendice:ecuaciones}) and approached numerically using the MATLAB\textsuperscript{\textregistered} suite for convenience. Several checks, like comparing with known theoretical solutions in simple vacuum cases where the boundary conditions are well defined, have been done prior to production calculations. For example, we have verified the fields of an infinite solenoid, or Eq.~\ref{ec:campoEthetav} for one coil. Incidentally, the results with several formulae for the conductivity have been also compared with a FORTRAN code \cite{zagorski20222-d-fluid-model} that uses a completely different numerical scheme, and with a finite-element-method code solving Eq. \ref{ec:lido19_4} in the same 2D geometry. In all cases the comparisons are satisfactory.

In the next sub-sections we describe a stepwise addition of ingredients in order to investigate their need to make the calculations approach the experimental information of input power and inductance in presence of the plasma. We recall that the input power is 50 kW and the inductance without plasma is about $9.6$ $\mu$H, decreasing very slightly in presence of the plasma, on the order of a few percent. According to Eq. \ref{ec:lido72_5}, the reduced inductance that corresponds to the void driver is $L^\mathrm{void}_\downarrow = 6.2$ $\mu$H. As mentioned in \ref{subsec:electrical}, we fix the amplitude of the RF coil current to $242$ A for  $I_\mathrm{PG}=0$ plasma profiles, and $218$ A for the $I_\mathrm{PG}=2.6$ kA case.

\subsection{Conductivity with local collisions and stochastic heating effects}

In order to start with a minimum amount of ingredients, several calculations have been dedicated to investigate at what extent the conductivity formula \ref{ec:conductividad_aprox} might explain the experimental information using only the local collisional processes, $\nu_\mathrm{local}$ (Eq.~\ref{ec:nu local}). The corresponding conductivities are well above a thousand S/m in all the plasma central part, and some hundred S/m near the Faraday shield lateral wall (plasma radius). This is due to the comparatively low values of $\nu_\mathrm{local}$, see figure \ref{fig:barridos_nu}. Using the $I_\mathrm{PG}=0$ profiles and the corresponding operational conditions of $I_\mathrm{RF}$, and neutral gas pressure and temperature, the iterative process diverges due to the excessive plasma response. Halving artificially the electron density, the ohmic power absorbed by the plasma is still above the nominal net power and the inductance decreases excessively, around a 50\%. This model for the conductivity is clearly inadequate for these plasmas.

  Adding the stochastic heating contribution through the equivalent collisionality, $\nu_\mathrm{st}$, takes the calculations closer to the expected values according to the experimental information. Figure \ref{fig:barridos_nu} shows that  $\nu_\mathrm{st} \gg \nu_\mathrm{local}$ in practically all relevant combinations of electron density and temperature. Therefore, we now include $\nu_\mathrm{st}$ to try variations in the most uncertain parameters in search for a possible match within experimental uncertainty.

\begin{figure}[htp] 
   \centering
   \includegraphics[width=0.85\columnwidth]{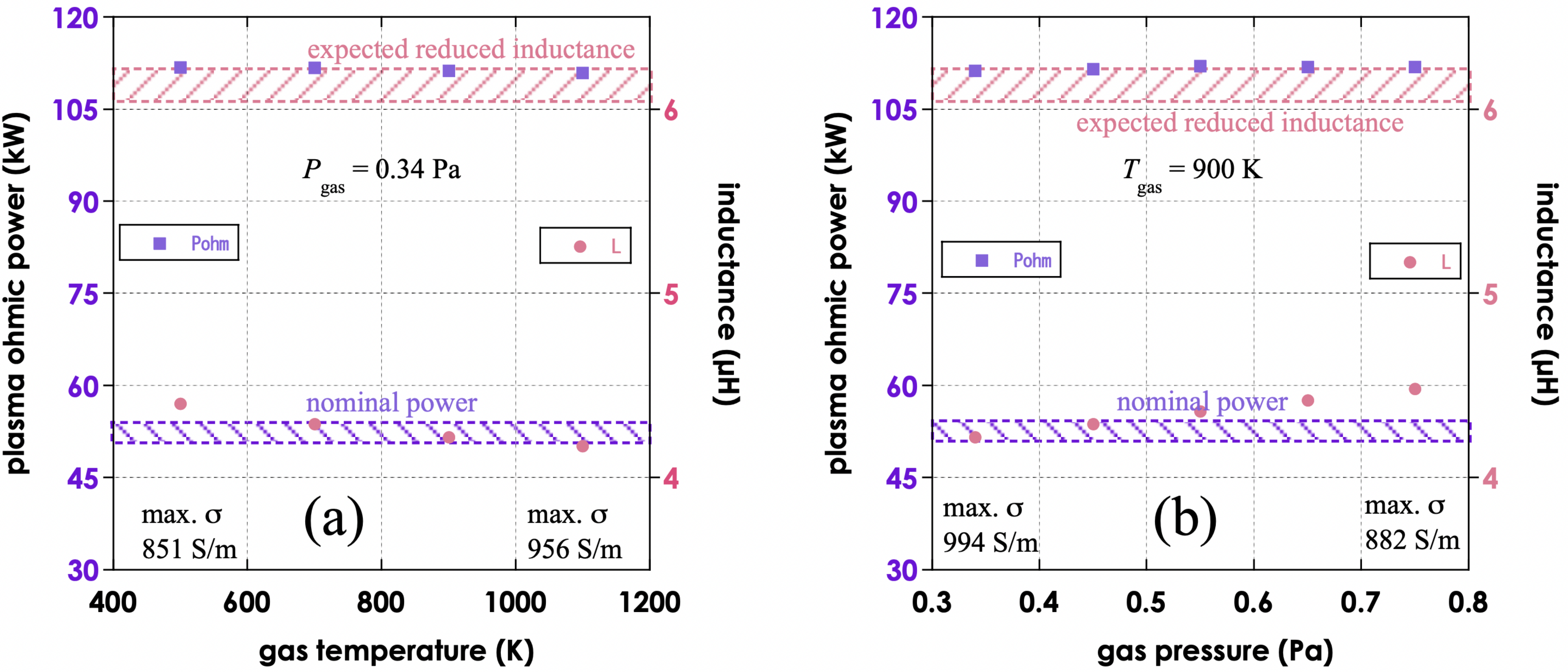} 
   \caption{Scans of the ohmic power and the reduced inductance, Eq. \ref{ec:lido72_5}, on the values of homogeneous gas temperature (a) and pressure (b) considering only local collisions and stochastic heating in the formulation of the plasma conductivity. The plasma profiles correspond to $I_\mathrm{PG}= 0$ in figures \ref{fig:olla68_1} and \ref{fig:ISPGbajoB}. The expected values are shown in shaded areas with corresponding colors.}
\label{fig:parameter scan only nu stochastic}
\end{figure}

Two global parameters that are subject to experimental uncertainty are the neutral gas pressure and temperature. 
 The gas pressure can decrease notably where the electron pressure is high. According to the measurements near the driver axis, we have an electron pressure $n_{e0}\mbox{[m}^{-3}\mbox{]}T_{e0}\mbox{[J]} \approx 1.8$ Pa, larger than the gas pressure $P_\mathrm{g} = 0.34$ Pa. Therefore, a considerable depletion of neutrals is expected in the center of the driver. For simplicity, we take a homogeneous neutral gas pressure but change its value as a whole in the scans. With respect to the temperature, we likewise consider a homogeneous value compatible with the experimental indications. Ongoing studies based on emission spectroscopy indicate that the gas temperatures, under reasonable assumptions, have a thermal component around 1000 K not strongly dependent on filter magnetic field or nominal power \cite{zaniol2020first-measureme}. 

The results of the neutral gas temperature and pressure scans are shown in figure \ref{fig:parameter scan only nu stochastic}.  
 We find that, with the conductivity Eq.~\ref{ec:conductividad_aprox} based on $\nu =\nu_\mathrm{local}+\nu_\mathrm{st}$, not only the values of ohmic power and driver inductance are far from the experimental ones for all values in the scans, but also the trends are opposite: when one parameter (e.g.~the gas pressure) makes the power tend --although very weakly-- towards the experimentally acceptable range, then the inductance moves away from it; and viceversa. The uncertainty in these two parameters, neutral gas pressure and temperature, cannot explain the mismatch between the experimental and calculated absorbed power and inductance.
 
\begin{figure}[htbp] 
   \centering
   \includegraphics[width=0.5\columnwidth]{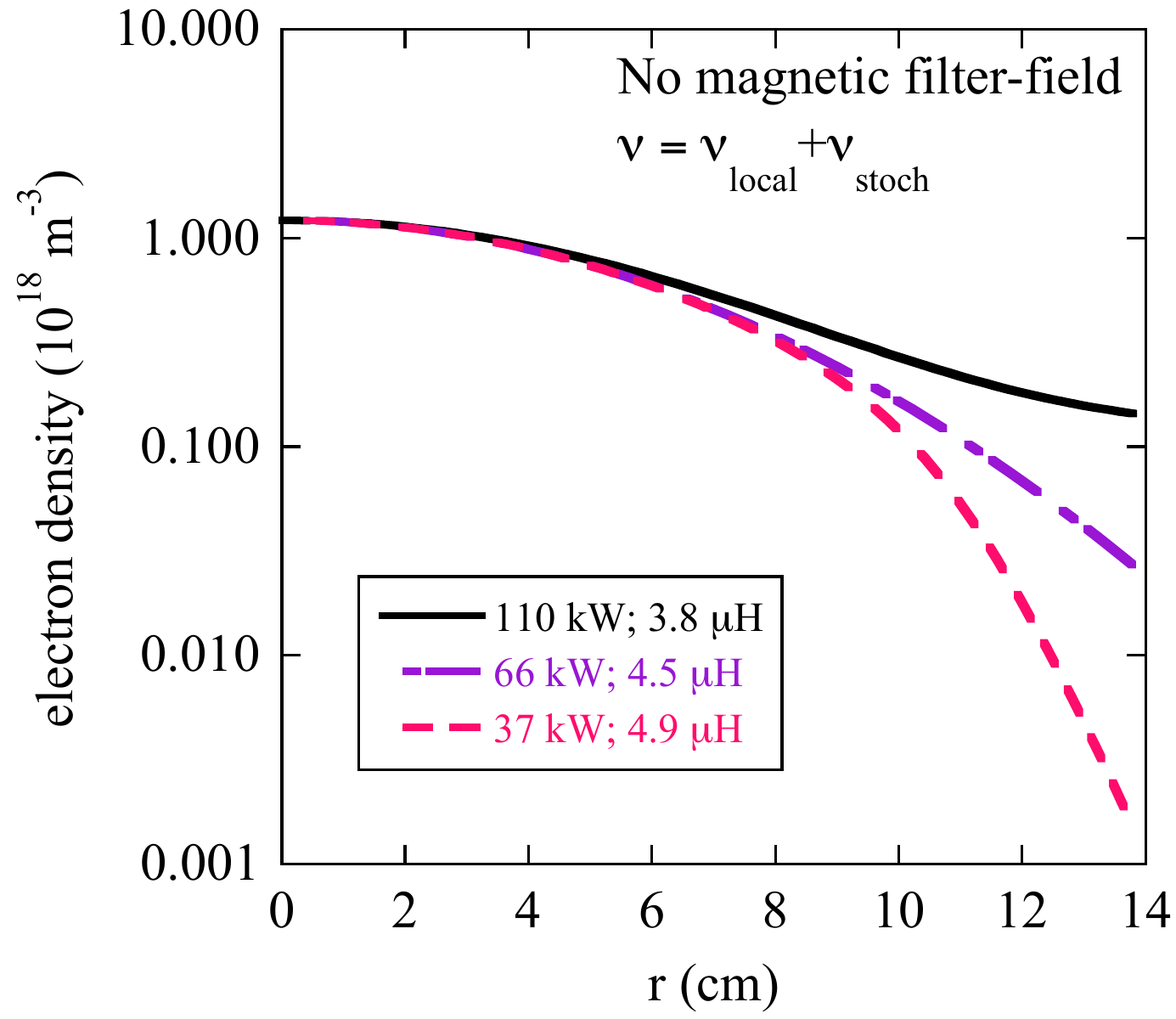} 
   \caption{Density profiles and corresponding values of plasma absorbed power (kW) and inductance ($\mu$H) using parameters of $I_\mathrm{PG}=0$ (no filter magnetic field) discharges. Reduced inductance without plasma: $L_\downarrow = 6.2$ $\mu$H.}
   \label{fig:ne_profile_scan}
\end{figure}

Another possibility within experimental uncertainty pertains the electron density and temperature profiles. According to emission spectroscopy data, the electron density and temperature in the outer half (in radius) should not be too different from the values used in the calculations of figure \ref{fig:parameter scan only nu stochastic}, respectively around $3\times 10^{17}$ m$^{-3}$ and $9$ eV. In the particular case of the electron density, its decrement near the plasma radius should be partly due to plasma compression despite the fact that the power remains at half the design value of $100$ kW per driver. Considering the possibility that the electron density near the border falls more than estimated with the radiative model, we have played with the profiles by changing the shape functions, Eq. \ref{ec:lido76_1}. Figure \ref{fig:ne_profile_scan} shows three profiles with decreasing edge values and the corresponding calculated values of power and reduced inductance. The ordinate is shown in logarythmic scale to highlight the differences. The delivered power to the plasma remains high in two of the cases, and the plasma inductance decreases too much in relation with the experiments. The latter is true also for the most favourable case where we have taken the density to almost negligible values near the plasma edge. 

\begin{figure}[htbp] 
   \centering
   \includegraphics[width=0.7\columnwidth]{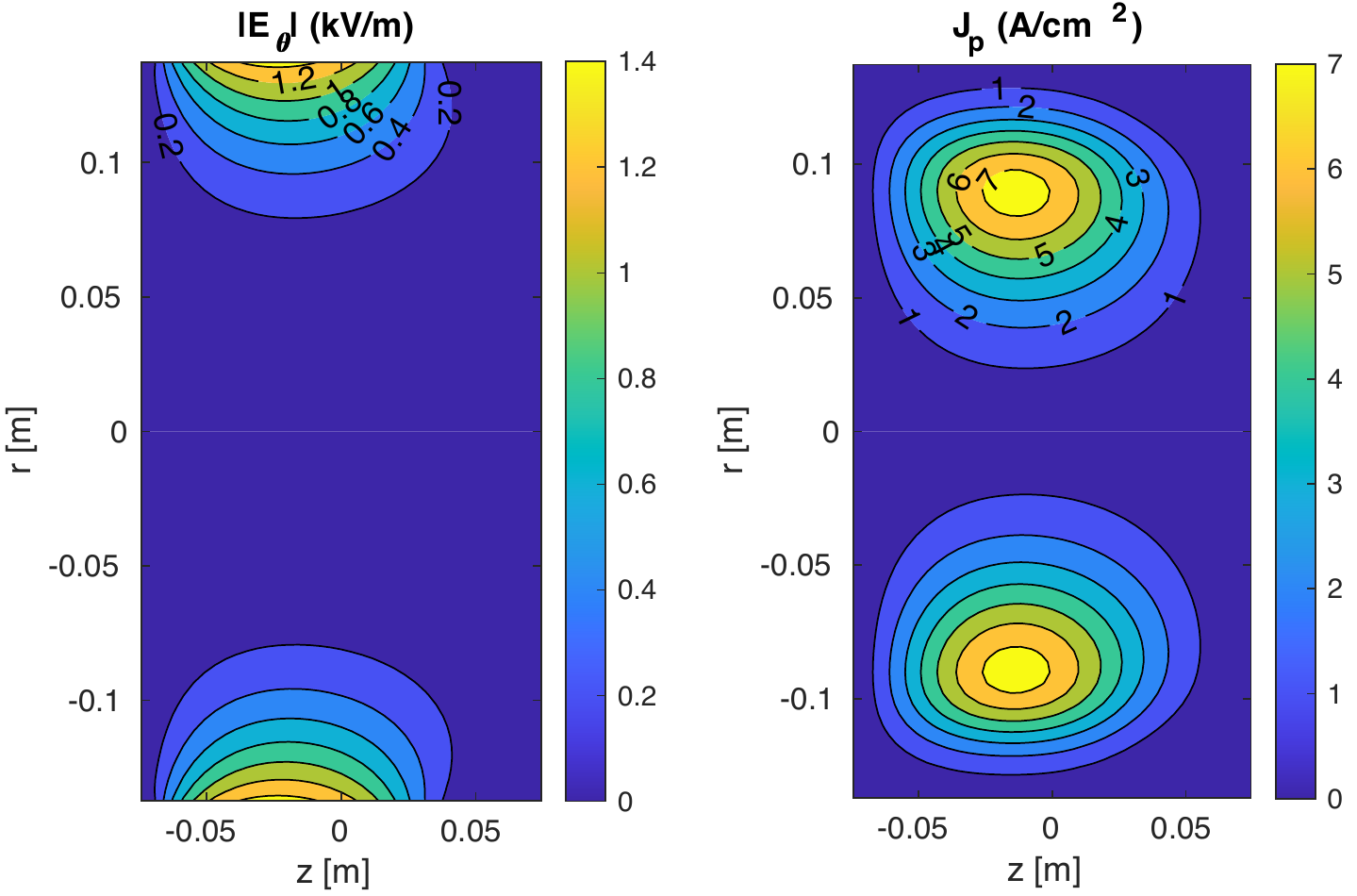} 
 \caption{2D distribution of the magnitude of the induced electric field, $|E_\theta |$, and the plasma current density, $J_\mathrm{p}$, for the case of $P_\mathrm{ohm} = 37$ kW in figure \ref{fig:ne_profile_scan}.}
   \label{fig:FF_casos027-028}
\end{figure}

  The results shown in figures \ref{fig:parameter scan only nu stochastic} and \ref{fig:ne_profile_scan} indicate that the inclusion of the stochastic formulation \cite{Vahedi1995Analytic-model-} gives conductivities still too large, resulting in ohmic powers generally above the nominal input power, $P_\mathrm{ohm} > P_\mathrm{RF}$.  The inductance falls always too short of the expected values in the sense that it decreases excessively from the void-driver value.  In essence, the reason is that the $\lesssim 50$ kW of absorbed power cannot be achieved unless the 2D distributions of induced electric field and plasma current density are far from overlapping. Figure \ref{fig:FF_casos027-028} shows respective contour maps of the amplitude of the induced electric field and current density for the favourable calculation of figure \ref{fig:ne_profile_scan}, i.e., the one yielding the smallest absorbed power. Since the conductivities grow rapidly with density, the current density peaks quite close to the RF coils where the induced electric field is still large. This provokes a large ohmic power density $\bb{J}\cdot \bb{E}$ and large net azimuthal currents, near 200 A in this case. The effect on the boundary conditions is large,  hence the decrement of the inductance. 
 
So far, 2D electromagnetic calculations suggest that the plasma current density must peak closer to the axis of the driver. Given the fact that the active currents are outside the plasma domain, this can only happen if the conductivities near the cylindrical side of the Faraday shield are still considerably reduced. Let us assume that the RF magnetic field can have a strong impact in the plasma conductivity due to the reduced mobility of the electrons. The amplitude of this field is, obviously, larger near the RF coil, which renders this contribution a good candidate to explain the experimental results in plasmas without static magnetic field. As mentioned in the Introduction, the problem of the electrical conductivity of RF plasmas is open to research and generally resolved numerically except in high-symmetry configurations. We have verified that reducing artificially the plasma conductivity in the regions of high $B=|\bb{B}|$ yields acceptable values of deposited power and inductance reduction. Therefore, we have taken a simple formulation of the $B$-field dependent electrical conductivity \cite{tuszewski1997inductive-elect} that plays this role.

\subsection{Inclusion of RF magnetic field effects}

When the RF magnetic field is included in the conductivity, Eq.~\ref{ec:ecuacion em} becomes non-linear. The azimuthal component of the electric field is linearly related with the same component of the vector potential in our problem. 
 In cylindrical coordinates we have two non-null components of the magnetic field,
$B_r = -\partial_z A_\theta $ and $B_z = (1/r) \partial_r (r A_\theta)$. The effect of the RF magnetic field on the plasma conductivity can be obtained following a procedure similar to the one leading to Eq.~\ref{ec:conductividad_aprox}, except that now the Lorentz force density is fully considered for the electron fluid. An estimate of the effect of the induced magnetic field on the plasma conductivity in cylindrical coordinates is provided in \cite{tuszewski1997inductive-elect}. Considering the limit $\omega^2 \ll \nu^2$, the expression is
\begin{equation}
\sigma_\mathrm{tus} = \sigma_\mathrm{dc} \left [  \frac{1}{(1+ \Omega_e^2/\nu^2)^{1/2}} - \imath \frac{\omega/\nu}{( 1 + \Omega_e^2/\nu^2)^{3/2}}  \right ],
\label{ec:sigma_tus}
\end{equation}
where $\Omega_e = eB/m_e$ and the ``direct current'' conductivity is $\sigma_\mathrm{dc} \equiv e^2 n_e/(m_e \nu)$. Note that, in terms of this definition, formula \ref{ec:conductividad_aprox} becomes
\begin{equation}
\sigma = \sigma_\mathrm{dc} \frac{\nu}{\nu + \imath \omega}
\label{ec:sigma_mis}
\end{equation}
and does not tend to Eq.~\ref{ec:sigma_tus} when $\Omega_e \to 0$, except in the mentioned limit $\omega^2 \ll \nu^2$. For our present purpose, we can use the model Eq.~\ref{ec:sigma_tus} even if it does no match Eq.~\ref{ec:sigma_mis} when the effect of the magnetic field is neglected.

\begin{figure}[htp] 
   \centering
   \includegraphics[width=0.95\columnwidth]{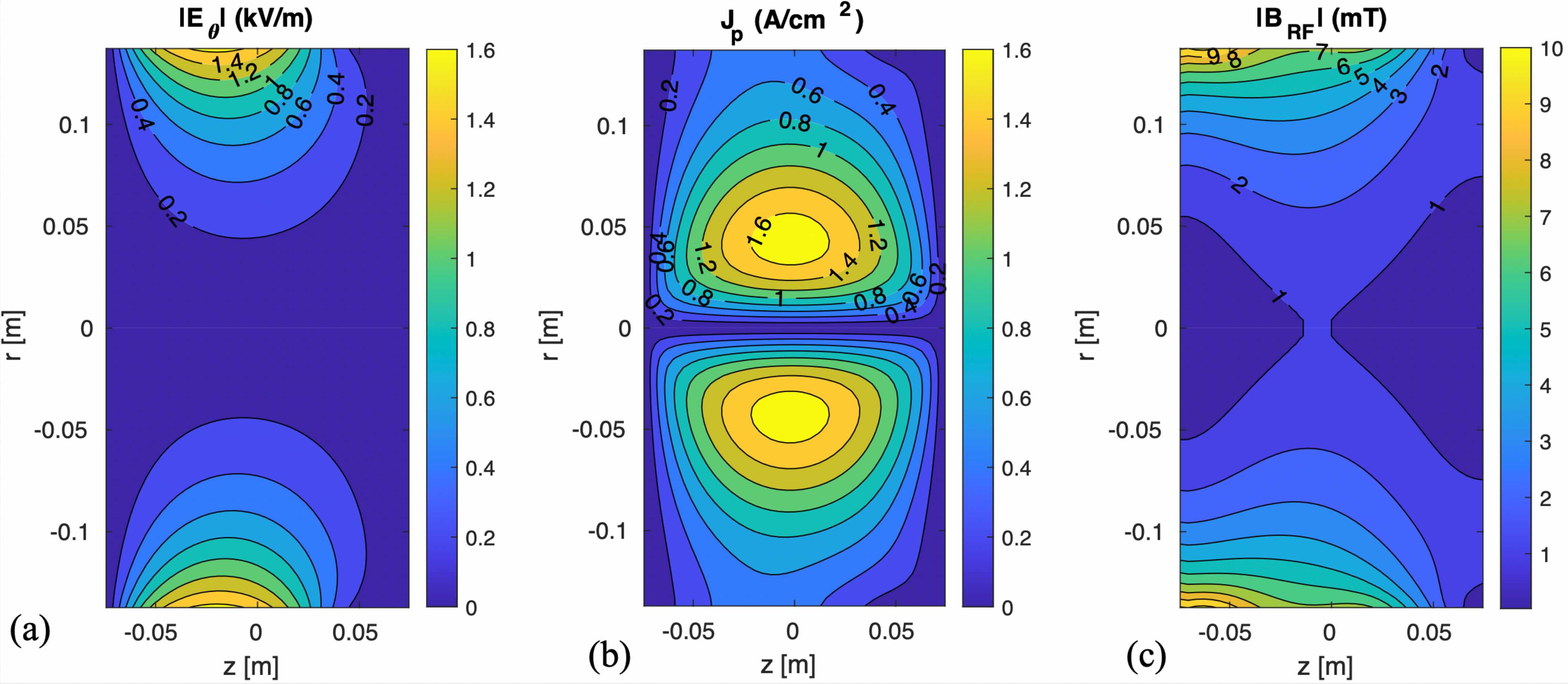} 
   \caption{Amplitude of the azimuthal induced electric field (a), current density (b) and modulus of the RF magnetic field (c) in a calculation with $I_\mathrm{PG}=0$ kA plasma profiles, see figures \ref{fig:olla68_1} and \ref{fig:ISPGbajoB}.}
\label{fig:cazuela008correcLb}
\end{figure}

 We have mentioned that the problem with a conductivity that depends on the induced electric field is non-linear, but the iterative procedure used to obtain the converged boundary conditions serves also to update the plasma conductivity with the successively calculated electric fields (hence vector potentials and magnetic fields).  Figure \ref{fig:cazuela008correcLb} shows the result on $I_\mathrm{PG}=0$ plasmas. In (a) we can see that the electric field penetrates more inside the plasma region than in the case without magnetic field effect (compare with figure \ref{fig:FF_casos027-028}), but the current density peaks much closer to the driver axis (b). In consequence, (i) the absorbed power falls to $P_\mathrm{ohm}= 20$ kW, well below the nominal $P_\mathrm{RF}=50$ kW; and (ii) the reduced inductance barely decreases, to $L = 6.16$ $\mu$H, less than 1\% of the corresponding void driver value. Both results are compatible with the experiments. Figure \ref{fig:cazuela008correcLb} (c) shows the magnitude of the RF magnetic field, where it can be appreciated that it reaches values $\approx 10$ mT near the RF winding. This is to be compared with the $\approx 3$ mT achievable inside the driver with the external static filter magnetic field. From the perspective of the model conductivity, the oscillating RF magnetic field alters the conductivity so as to become the dominant effect in our plasma conditions.

\subsection{Inclusion of a static magnetic field}

We present here a calculation based on plasma parameters obtained with the application of the filter magnetic field, $B_\mathrm{f}$. This is intended to be the main mode of operation of the sources of SPIDER. The electron density and temperature distributions are based on experimental data as before, and they are set in the calculations via Eq.~\ref{ec:lido76_2}, that is, with profiles corresponding to $I_\mathrm{PG}=2.6$ kA discharges. This current is close to the maximum available in the plasma grid during the 2020 campaigns, and therefore represents about the maximum static magnetic field attainable in the present experiments.
 Following previous works \cite{Recchia2021studies-on-powe}, we include the static magnetic field using the same formulation for the conductivity in presence of the RF magnetic field, Eq. \ref{ec:sigma_tus}, but now substituting  the cyclotron frequency by an ``equivalent'' value such that $\Omega_\mathrm{eq}^2 = \Omega_e^2 + \Omega_\mathrm{f}^2$, where $\Omega_\mathrm{f} = eB_\mathrm{f}/m_e$ is the cyclotron frequency that corresponds to the static filter magnetic field \cite{cavenago2012models-of-radio}. Note that this is also a simplified formulation where only the modulus, not the vectorial character, of the magnetic field is taken into account. The consideration of an order-two tensorial conductivity would require three-dimensional calculations.

The filter magnetic field is approximately perpendicular to the axis of the drivers \cite{marconato2021an-optimized-an} and its modulus increases from the back side towards the opening to the expansion region. The field is similar, but not identical, in all drivers due to the obliged proximity of the circuitry of the driving current $I_\mathrm{PG}$. A representative expression for the filter magnetic field magnitude, depending on its maximum $B_0$, can be taken as
\begin{equation}
B_\mathrm{st}(r,z) = \frac{B_0 - B_{0a}}{2} \left ( 1-\tanh \frac{\delta_B-z}{\delta_B} \right ) + B_{0a},
\label{ec:olla62_3}
\end{equation}
where the offset $B_{0a}$ and the width $\delta_B$ are chosen to approximate the experimental data. Here we take $B_{0a} = 1.8$ mT and $\delta_B=0.08$ m for a magnetic field $B_\mathrm{f} = 4$ mT, which gives a maximum of $3.6$ mT at the opening to the expansion region.

\begin{figure}[htbp] 
   \centering
   \includegraphics[width=0.6\columnwidth]{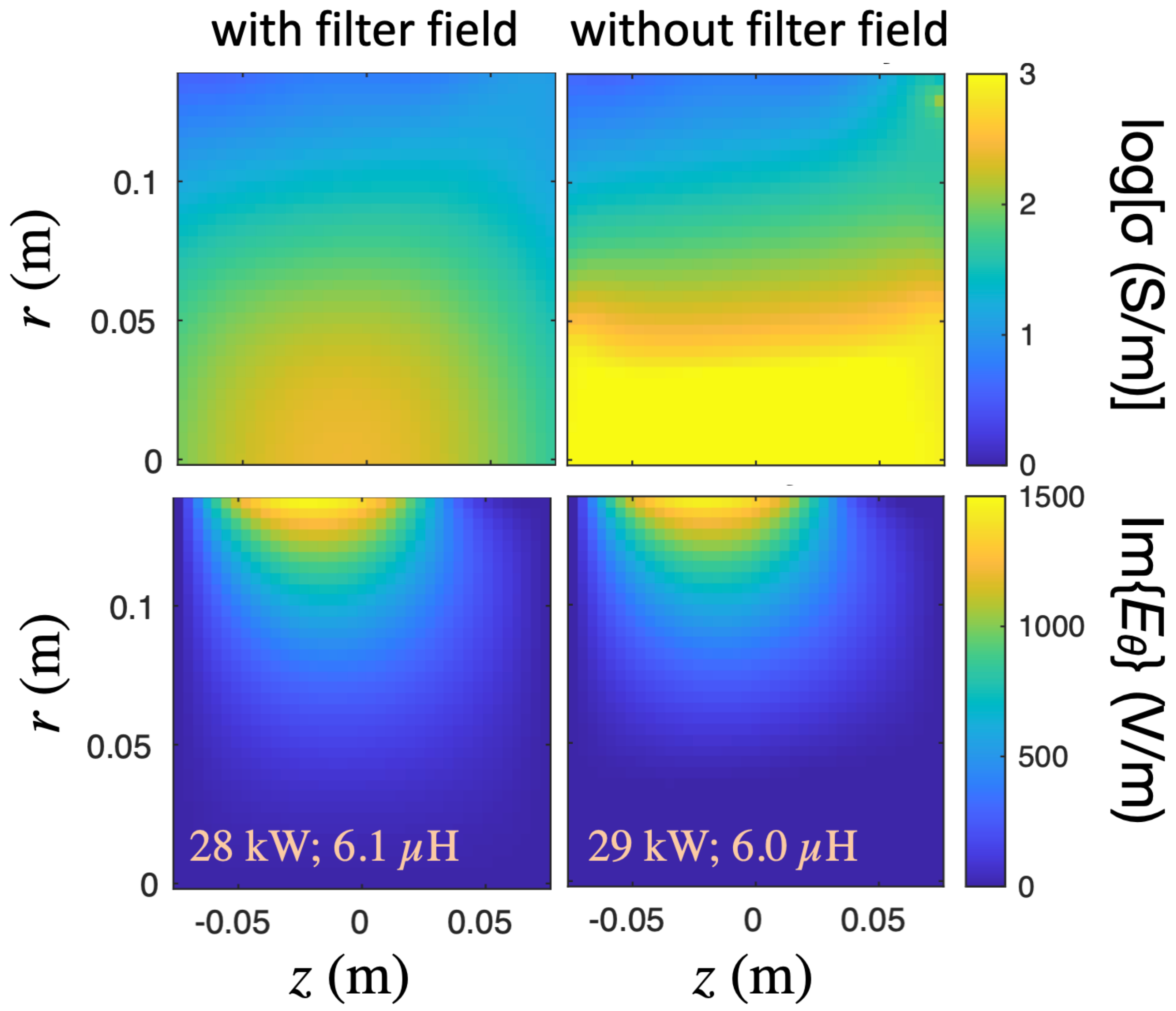} 
 \caption{2D distributions of plasma conductivity (top) and imaginary part of the induced electric field $E_\theta$ (bottom) based on  plasma data from SPIDER discharges with $I_\mathrm{PG}=2.6$ kA,  including the effect of the static filter magnetic field (left) or excluding it (right).}
 \label{fig:sigmas_ImEq}
\end{figure}

In figure \ref{fig:sigmas_ImEq} we show the 2D maps of the conductivity (top) and the imaginary part of the induced electric field (bottom) for two calculations based on the $I_\mathrm{PG}=2.6$ kA plasma profiles. Both calculations include the effect of the RF magnetic field on the conductivity, but the static magnetic filter-field is only considered in the left panels, as indicated. Since the models used take into account only the magnitudes of the magnetic field, the effect of the filter magnetic field on the absorbed power is rather weak once the RF field has been considered. Indeed, even though the plasma conductivity increases considerably near the driver axis when the static filter-field is not considered, the induced electric field (only the dominant out-of-phase part is shown) remains practically unchanged.

\section{\label{sec:discusion}Discussion}

Section \ref{sec:resultados} has been devoted to justify the main elements that yield acceptable estimates of absorbed power and decrement in driver inductance with plasma in conditions of SPIDER discharges. Using experimental data from discharges \emph{without} filter magnetic field, we have found that there is some essential ingredient in the plasma conductivity that should reduce it mainly where the oscillating RF magnetic field is larger. We have reproduced this behaviour with the simple formulation Eq.~\ref{ec:sigma_tus}. We have also tried a formulation for the static filter magnetic field (figure \ref{fig:sigmas_ImEq}). Here we find a weak additional effect because the conductivity is reduced mainly where the induced electric field is already very low.  However, from the experimental viewpoint it is clear that the filter magnetic field makes a notable difference in plasma parameters. Therefore, the static field must be considered in transport calculations. 
 In the calculations that follow we retain all the ingredients tested so far. 

\begin{figure}[htbp] 
   \centering
   \includegraphics[width=0.9\columnwidth]{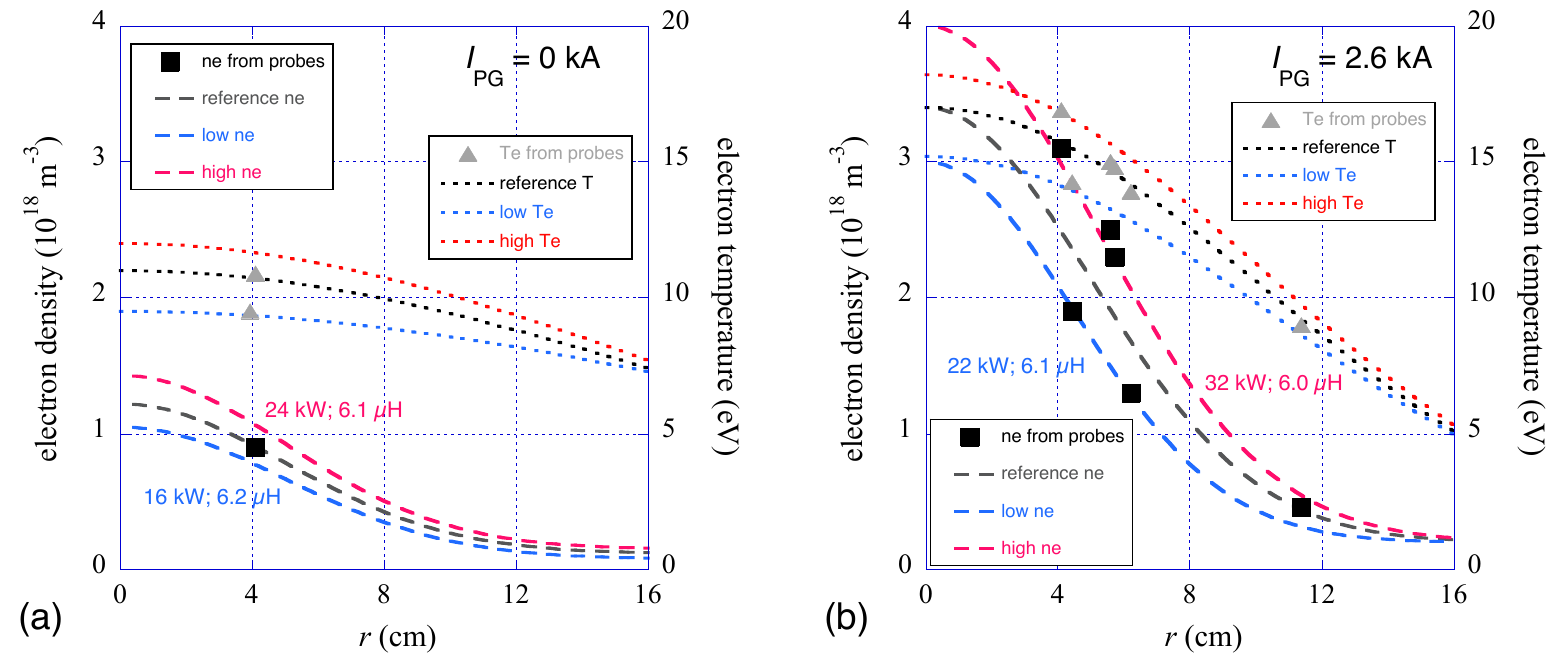} 
   \caption{Extreme electron density (dashed lines) and temperature (dotted lines) profiles at $z=0$ used in calculations without (a) and with (b) filter magnetic field. The corresponding calculated values of absorbed power and reduced inductance are indicated for the low (blue) and high (red) profiles. Experimental data are shown with squares for the density and triangles for the temperature. The reference profiles Eq. \ref{ec:lido76_2}--\ref{ec:lido76_1} are shown with black lines.}
   \label{fig:profiles_ISPG}
\end{figure}

The deposited power is sensitive to the electron density and temperature values around mid-radius, where the induced electric field is still intense (here the reduced conductivity associated to the RF magnetic field is responsible for the large field penetration) and the plasma conductivity starts having large values so as to provoke intense current densities. For this reason, we present the final results giving margins according to the limit profiles presented in figure \ref{fig:profiles_ISPG}, where the radial functions of the reference profiles for each case, without and with magnetic filter field, are shown with black discontinuous lines and the probe data are shown with symbols. Since the uncertainty of data is larger in the radial dimension, we have performed several calculations changing the form functions for the radial functions in Eqs.~\ref{ec:lido76_2} and \ref{ec:lido76_1}, as shown in the figure. The purpose is to inform on the sensibility of the assumed model conductivity to the experimental uncertainty. The values of absorbed power and inductance are labelled in corresponding colours for the extreme profiles considered in this study. The power remains, in all cases, below the nominal $P_\mathrm{RF}=50$ kW per driver. The inductances decrease less than a 3\%, also in agreement with the experimental findings \cite{jain2022investigation-o}, except for the 32 kW case with $I_\mathrm{PG}=2.6$ kA.  
 The obtained transfer efficiencies range, in the $I_\mathrm{PG}=0$ kA case, from $P_\mathrm{ohm}/P_\mathrm{RF} = 16 \mbox{ kW}/50\mbox{ kW} \approx 35\%$ to $24\mbox{ kW}/50\mbox{ kW} \approx 50\%$; while in the $I_\mathrm{PG}=2.6$ kA case, they range approximately from $45\%$ to $65\%$. As mentioned, however, this latter case might be overestimated because the inductance is expected to decrease somewhat less. Globally, we find power transfer efficiencies around a 50\%, apparently higher in presence of the filter magnetic field. 

It is worth making a comment on the power transfer efficiency. We observe that the entire problem scales with $I_\mathrm{RF}$ through the boundary conditions (Eq.~\ref{ec:campoEthetav}), and the calculated absorbed power changes with $I_\mathrm{RF}^2$.  Based on the equivalent resistances found in \cite{Recchia2021studies-on-powe} we could allow for a change of, say, a 10\% with respect to the values we have set in order to obtain $I_\mathrm{RF}$. Let us call $R_\mathrm{eq}$ our choice of equivalent resistance and let $R_{\mathrm{eq},1}$ be a 10\% different equivalent resistance giving rise to a new current $I_{\mathrm{RF},1}$. For a fixed nominal power, the ratio $I_{\mathrm{RF}}^2/I_{\mathrm{RF},1}^2 = R_{\mathrm{eq},1}/R_{\mathrm{eq}}$, which would give the same ratio among absorbed powers. The range of calculated powers in figure \ref{fig:profiles_ISPG} could then be shifted up or down on the order of the same percentage, too little to make a considerable difference with respect to the conclusions about the importance of the RF magnetic field in the conductivity of the plasmas in the SPIDER sources. 
 
Finally, we would like to underline the fact that the calculations here presented, based on experimental data, do not pretend to prove the validity of the conductivity models, but just give a first approximation to the necessary ingredients. Indeed, the model for the stochastic conductivity is based on the evaluation of a skin depth that is not necessarily consistent with the evaluated penetration of the electric field; and the penetration of the induced electric field is a consequence of the smaller conductivity provoked by the RF magnetic field (Eq.~\ref{ec:sigma_tus}), but this should be taken as a practical formulation rather than a physical model \cite{smolyakov2000on-nonlinear-ef}.
 The same applies to the model for the static filter magnetic field \cite{cavenago2012models-of-radio}, which considers the modulus while the direction of the field is obviously important. With these considerations in mind, the
 present models for the 2D electromagnetic calculations in the drivers of the SPIDER device give acceptable values of the ICP  deposited power and current density distributions, which makes them suitable for a first formulation to be considered in a 2D transport code.  Even if it is acknowledged that the conductivity models for the magnetic field effects can be questioned from the physics perspective (like the 2D restriction of the calculations) and detailed modelling is probably mandatory \cite{chen2006nonlinear-effec}, any future formulation of the electrical conductivity should not give too different numbers from those here found because we are using plasma experimental data as input.  At his respect, the present work points to RF magnetic field effects on the electrical conductivity of the plasma as a necessary ingredient in the physics of ICP in SPIDER discharges.
 
 Further improvements in calculations for SPIDER plasma sources (and, correspondingly, for the ITER device) should also benefit from ongoing self-consistent calculations (RF inductive heating plus transport) \cite{zagorski20222-d-fluid-model,zagorski20232d-simulations}. There is also ongoing research to extend the present calculations to three dimensions, including non axi-symmetric components like the Faraday shield \cite{lopezbruna2022Three-dimensional}. Such 3D electromagnetic calculations of the induction process might be used to investigate more detailed models, for instance including anisotropic or non local models of the electrical conductivity. This would be interesting to check at what extent the present 2D calculations can be considered as a practical approximation of the plasma-RF coupling in other applications.

\section{\label{sec:conclusion}Conclusion}

This work documents the first electromagnetic simulations of the coupling between the RF field and the plasma inside the drivers of the SPIDER device. Initial boundary conditions are set so as to reproduce the boundary vacuum fields and, consequently, the experimental driver inductance without plasma. An iterative process provides the reduction of the driver inductance in presence of the plasma. Using experimental information about the electron density and temperature inside the drivers, the calculated heating power and the reduced inductance can be compared with their experimental counterparts depending on the formulation provided for the plasma conductivity. In this way, it is found that local collisional processes and, likely, non-local processes associated to stochastic heating, cannot explain the experimental conditions. The effect of  the magnetic field is taken from simple analytical formulations that use only its modulus. Despite their simplicity and questionable physics, they happen to make the 2D electromagnetic calculations compatible with the experimental knowledge of input power and inductance reduction with plasma.  Therefore, the RF magnetic field is posed as a necessary ingredient to understand the ICP in discharges created without and with static filter magnetic field in the SPIDER device. Aside from physical interpretations, this paper informs on what kind of change in the electrical conductivity is necessary in our plasmas so as to acceptably approach the experimental values. From this perspective, the present 2D electromagnetic calculations confirm the suitability of the conductivity models used in previous studies \cite{Recchia2021studies-on-powe}, and can be taken as a practical starting tool in 2D fluid-transport codes for this device \cite{zagorski20222-d-fluid-model,zagorski20232d-simulations}.

\section*{Acknowledgement}

This work has been carried out within the framework of the EUROfusion Consortium, funded by the European Union via the Euratom Research and Training Programme (Grant Agreement No 101052200 — EUROfusion). Views and opinions expressed are however those of the author(s) only and do not necessarily reflect those of the European Union or the European Commission. Neither the European Union nor the European Commission can be held responsible for them. This work has been carried out within the framework of the ITER-RFX Neutral Beam Testing Facility (NBTF) Agreement and has received funding from the ITER Organization. The views and opinions expressed herein do not necessarily reflect those of the ITER Organization.

\appendix

\section{\label{apendice:ecuaciones}Induction equation}

In general, the vortex source density of the magnetic field is
$$
\nabla \times \mathbf{B} = \mu_0 \mathbf{J} + \mu_0 \partial_t (\epsilon \mathbf{E}),
$$
where $\mathbf{J}=\mathbf{J}_\mathrm{b}+\mathbf{J}_\mathrm{p}$ are the material current densities, respectively in the windings and in the plasma; and $\epsilon(\mathbf{r},t)$ is the permittivity that we take as a scalar function in order to simplify. Disregarding the electrical circuits that feed the RF winding, we associate $\mathbf{J}_\mathrm{b}$ to a filamentary current in the winding. We take the time dependence of this current as an ideal harmonic
\begin{equation}
I_\mathrm{RF} = \Re \{ I_\mathrm{RF} e^{\imath \omega t} \}
\label{ec:lido18_5}
\end{equation}
where $I_\mathrm{RF}$ is real, and the winding as a set of circular loops encircling the driver.

From the curl of the induction law, $\partial_t \nabla \times \mathbf{B} = - \nabla \times \nabla  \times \mathbf{E} =  \nabla^2 \mathbf{E}- \nabla (\nabla \cdot \mathbf{E})$, one obtains a vector differential equation
\begin{equation}
\nabla^2 \mathbf{E} - \nabla \left ( \frac{\rho}{\epsilon} \right )  = \mu_0 \partial_t \mathbf{J} + \mu_0 \partial_{tt} (\epsilon \mathbf{E}).
\label{ec:lido16_9}
\end{equation}
Here we have substituted the divergence of $\bb{E}$, so $\rho$ is the charge density. We are interested in spatial scales much larger than the Debye length and assume, in addition, negligible capacitive coupling. In these conditions we can drop the term $\nabla (\rho/\epsilon)$.

We exploit the cylindrical symmetry of the device to simplify the system of equations. We remind here the cylindrical expression for the Laplacian of a vector,
$$
\nabla^2 \mathbf{E} = 
\left(
\begin{array}{c}
\nabla^2 E_r - \frac{2}{r^2}\partial_\theta E_\theta - \frac{E_r}{r^2}  \\
\nabla^2 E_\theta + \frac{2}{r^2}\partial_\theta E_r - \frac{E_\theta}{r^2}  \\
\nabla^2 E_z
\end{array}
\right),
$$
and of a scalar function
$$
\nabla^2 E_j = \frac{1}{r} \partial_r (r \partial_r E_j) +  \frac{1}{r^2} \partial_{\theta\theta} E_j + \partial_{zz} E_j.
$$
Let us assume beforehand that there is indeed cylindrical symmetry. Then there are no dependencies on $\theta$ and the three equations involved in Eq.~\ref{ec:lido16_9} reduce to one equation for the only component of the electric field,
\begin{equation}
\frac{1}{r}\partial_r (r \partial_r E_\theta ) + \frac{1}{r^2}\partial_{\theta\theta} E_\theta + \partial_{zz}E_\theta - \frac{E_\theta}{r^2} - \mu_0 \partial_{tt}(\epsilon E_\theta) = \mu_0 \partial_t J_\theta
\label{ec:lido17_5}
\end{equation}

A rapid dimensional analysis shows that, due to the relatively low angular frequency $\omega$, the displacement currents can be neglected in the plasma region, $\partial_{tt}\to 0$. Therefore, associating the plasma response $J_{\mathrm{p}\theta}$ to the induced electric field via a scalar conductivity,
$$
J_\theta = J_{\mathrm{b}\theta} + J_{\mathrm{p}\theta} = J_{\mathrm{b}\theta} + \sigma E_{\theta},
$$
the expression \ref{ec:lido17_5} reduces to 
\begin{equation}
\frac{1}{r}\partial_r (r \partial_r E_\theta ) + \partial_{zz}E_\theta - \frac{E_\theta}{r^2} - \mu_0 \partial_{t}(\sigma E_\theta) = \mu_0 \partial_t J_{\mathrm{b}\theta}.
\label{ec:lido18_4}
\end{equation}
Recalling Eq.~\ref{ec:lido18_5}, the solution can take the form
\begin{equation}
E_\theta (r,z;t)=\tilde{E}(r,z) e^{\imath \omega t}
\label{ec:lido31_2}
\end{equation}
with a complex spatial part, $\tilde{E} \in \mathbb{C}$, which is the component that the code must solve for. The imaginary part of $\tilde{E}$ explains the (position dependent) phase of the azimuthal electric field with respect to the current in the RF coil. Writing the solution as $\tilde{E}=|\tilde{E}|e^{\imath \varphi}$, we have
$$
E_\theta (r,z;t)=|\tilde{E}(r,z)| e^{\imath [\omega t + \varphi (r,z)]}
$$
and the real field amplitude is $| \tilde{E} |$.

The differential equation \ref{ec:lido18_4}, once discretized using some numerical differences method, becomes a linear system of equations on the discretized variable $E_\theta$. Thus, the equation 
\begin{equation}
\partial_r (r \partial_r E_\theta )  - \frac{E_\theta}{r}  + r \partial_{zz} E_\theta - \imath \omega r \mu_0 \sigma E_\theta = \mu_0 r \partial_t J_{\mathrm{b}\theta}
\label{ec:lido19_4}
\end{equation}
becomes the linear system
$$
\mathsf{A}\cdot \mathbf{E} = \mathbf{b},
$$
where the coefficients $A_i^j$ of $\mathsf{A}$ depend on the numerical stencil for the Laplacian and the induced currents ($\propto E_{\theta j}$), while the elements $b_j$ of $\mathbf{b}$ are related with the boundary conditions. In our case we solve the homogeneous Eq.~\ref{ec:lido19_4} because the calculation domain does not include the external currents $J_{\mathrm{b}\theta}$, but the boundary conditions depend on them and impose fixed values that move from the left-hand-side to $\mathbf{b}$.

\bibliographystyle{unsrt} 
\bibliography{elmag2D}

\end{document}